\documentclass{aa}
\pdfoutput=1
\usepackage[varg]{txfonts}
\usepackage[colorlinks=true,linkcolor=blue,citecolor=blue]{hyperref}
\usepackage{graphicx}

\begin{document} 

 \title{Redundant apodization for direct imaging of exoplanets I: Robustness to primary mirror segmentation-induced errors \\
 outside the segment diffraction limit}
 \titlerunning{Redundant apodization for direct imaging of exoplanets I}
\authorrunning{Leboulleux et al.}
  \author{Lucie Leboulleux \inst{1}, Alexis Carlotti \inst{1}, Mamadou N'Diaye \inst{2}}
  \institute{Univ. Grenoble Alpes, CNRS, IPAG, 38000 Grenoble, France
  \and Université Côte d’Azur, Observatoire de la Côte d’Azur, CNRS, Laboratoire Lagrange, France
  \\
             \email{lucie.leboulleux@univ-grenoble-alpes.fr}} 

  \abstract
  % context heading (optional)
   {Direct imaging and spectroscopy of Earth-like planets and young Jupiters require contrast values up to $10^{6}-10^{10}$ at angular separations of a few dozen milliarcseconds. To achieve this goal, one of the most promising approaches consists of using large segmented primary mirror telescopes with coronagraphic instruments. While they are able to reach high contrast at small angular separations, coronagraphs are highly sensitive to wavefront errors, however. The segmentation itself is responsible for phasing errors and segment-level vibrations that have to be controlled at a subnanometric accuracy.}
  % aims heading (mandatory)
   {We propose an innovative method for a coronagraph design that allows a consequent relaxation of the segment phasing and stability constraints for low segment-count mirrors and generates an instrument that is more robust to segment-level wavefront errors.}
  % methods heading (mandatory)
   {This method is based on an optimization of the coronagraph design that includes a segment-level apodization. This is repeated over the pupil to match the segmentation redundancy and improves the contrast stability beyond the minimum separation set by the single-segment diffraction limit.}
  % results heading (mandatory)
   {We validate this method on a Giant Magellan Telescope (GMT)-like pupil (seven circular segments) for two coronagraph types: apodized pupil Lyot coronagraphs, and apodizing phase plate coronagraphs. For the apodized pupil Lyot coronagraphs, redundant apodization enables releasing the piston phasing constraints by a factor of 5 to 20 compared to classical designs. For the apodizing phase plate coronagraphs, the contrast remains almost constant up to 1 radian RMS of the phasing errors. We also show that redundant apodizations increase the robustness of the coronagraph to segment tip-tilt errors, as well as to missing segment errors.}
   %Conclusion:
   {Redundant apodization enables reducing or even removing any constraints on the primary mirror segment phasing at the price of larger angular separations and lower throughputs. This method cannot be applied to higher-segment count mirrors such as the ELT or the TMT, but it is particularly suitable for low segment-count mirrors (fewer than $\sim 20$ segments) such as the GMT aperture. These mirrors aim for high-contrast imaging of debris disks or exoplanets down to 100 mas.}
    

   \maketitle

%%%%%%%%%%%%%%%%%%%%%%%%%%%%%%%%%%%%%%%%%%%%%%%%%%%%%%%%
\section{Introduction}
\label{s:Introduction}
%%%%%%%%%%%%%%%%%%%%%%%%%%%%%%%%%%%%%%%%%%%%%%%%%%%%%%%%

The next generation of spectro-imagers installed on giant telescopes will give access to smaller angular separations and fainter planets with the aim to characterize young Jupiters and the very first exo-Earths: typically, they target contrasts down to $10^{-8}$ at angular separations smaller than $0.1$'' and are combined with spectrometers or integral field spectrographs. In the coming years, the current high-contrast imagers on 8-meter class ground-based telescopes should be upgraded to access smaller angular separations and deeper contrasts than they currently do: the Very Large Telescope (VLT)/SPHERE+ (Spectro Polarimetric High contrast Exoplanet REsearch) for 2025 \citep{Boccaletti2020}, Gemini/GPI2.0 (Gemini Planet Imager) \citep{Chilcote2020}, and Subaru/SCExAO (Subaru Coronagraphic Extreme Adaptive Optics) \citep{Lozi2020}. In addition, various instruments with high-contrast capability are being developed for the coming 20 years. On the front line, we find the first light spectro-imagers of the giant ground-based telescopes, such as the first-light spectro-imagers of the European Extremely Large Telescope (ELT, $\sim$ 2027) Mid-Infrared E-ELT Imager and Spectrograph (METIS) \citep{Kenworthy2016, Brandl2018}, and High Angular Resolution Monolithic Optical and Near-infrared Integral field spectrograph (HARMONI) \citep{Carlotti2019}, or the Infrared Imaging Spectrograph (IRIS) for the Thirty Meter Telescope (TMT) \citep{Larkin2016}. Ground-based second-generation imagers such as Planetary Camera and Spectrograph (PCS) for the ELT (2035-2040) \citep{Kasper2013}, Planetary Systems Imager (PSI) for the TMT \citep{Fitzgerald2019}, GMagAO-X for the Giant Magellan Telescope (GMT) \citep{Males2018} will target fainter planets closer to their host star down to young Jupiters and telluric planets. The recent NASEM Astro2020 decadal report \citep{NASEM2021} also recommends the development of a large space telescope to observe planets down to Earth-like planets, such as the Large UV-Optical-IR (LUVOIR) B \citep{TheLUVOIRTeam2019} and Habitable Exoplanet (HabEx) \citep{Gaudi2020} candidates.

This effort has already started and is ongoing, with the improvement and development of new tools to equip the instruments and telescopes: in particular, the primary mirrors are segmented to reach larger diameters \citep{Yaitskova2002, Yaitskova2003}, coronagraphs able to provide very high contrast are becoming robust to amplitude and phase aberrations \citep[e.g., ][]{Carlotti2011, Carlotti2014, N'Diaye2016, Fogarty2018, Ruane2018}, and wavefront sensors and controllers that are compatible with coronagraphs and that are sensitive enough to correct for very faint aberrations are being developed \citep{Jovanovic2018}. For large phasing aberrations, the ELASTIC algorithm \citep{Vievard2017}, the wavelength sweep \citep{Cheffot2020}, the COFFEE estimator \citep{Paul2013, Leboulleux2020}, and Zernike wavefront sensors \citep{Dohlen2006, Surdej2010, Vigan2011} have been tested successfully. In particular, the Zernike wavefront sensor called ZELDA can align segments down to a few nanometer root mean square (RMS) \citep{N'Diaye2016a, Janin-Potiron2017}. Similarly, the LAPD algorithm \citep{Vievard2020}, the Mach-Zehnder sensor \citep{Yaitskova2004}, and the self-coherent camera can in optimal cases even reduce the aberrations to the subnanometer level \citep{Janin-Potiron2016}. Mechanical sensors also provide a high accuracy for segment phasing \citep{MacMartin2004}, and we can cite the promising work of \cite{Saif2017}, who  detected phasing errors down to a dozen picometers.

This phasing-error detection and correction is a key component of high-contrast imagers because the coronagraphic performance is highly sensitive to phase errors and pupil discontinuities. For instance, the tolerancing of Earth-like planets imagers has established drastic constraints of about $10$ pm RMS for the segment alignment over 10 minutes to obtain a $10^{-10}$ contrast with the LUVOIR telescope \citep{Stahl2015, Stahl2016, Stahl2017, Stahl2020, Leboulleux2017, Leboulleux2018, Leboulleux2018a, Laginja2019, Laginja2020, Laginja2021, TheLUVOIRTeam2019}. The ELT-HARMONI instrument is being designed to achieve a more moderate $10^{-6}$ contrast after post-processing, which requires the noncommon path aberrations to be below $10-15$ nm RMS at any given time.

Current developments in coronagraphy and wavefront control involve robustness to the mirror segmentation and to the spiders. For instance, apodized pupil Lyot coronagraphs (APLCs) \citep[e.g., ][]{Soummer2009, N'Diaye2015, Zimmerman2016}, vortex coronagraphs \citep[e.g., ][]{Ruane2015, Ruane2015a}, phase-induced amplitude apodization (PIAA) coronagraphs \citep[e.g., ][]{Guyon2014}, and the active compensation of aperture discontinuities (ACAD-OSM) \citep{Pueyo2013, Mazoyer2018, Mazoyer2018a} are now optimized to compensate for these effects. However, these solutions correct for amplitude errors (aperture shape, spiders, and missing segments) and should be combined with a wavefront control system to handle dynamic phase errors.

In this paper, we propose an innovative method to design coronagraphs that are robust to segment misalignments and instabilities outside the single-segment diffraction limit. It derives from the pair-based analytical model for segmented telescope imaging from space (PASTIS) \citep{Leboulleux2018, Laginja2021} and consists of an apodization of the segment to reduce its impact on the final image. This technology can be compared to the shaped pupils \citep{Vanderbei2003, Kasdin2003, Carlotti2011} in the amplitude apodization case and the apodizing phase plate coronagraphs (APPs) in the phase apodization case \citep{Codona2007, Kenworthy2010, Carlotti2013, Por2017}. This segment apodization is then repeated over the pupil to mimic the primary mirror segmentation redundancy, creating a so-called redundant apodized pupil (RAP). This RAP can be combined with a second coronagraphic design step to highten the contrast in the high-contrast region in the coronagraphic image of the star even more (called the dark zone), such as classical APLCs. Whether it is combined with this second design step or not, this RAP design reduces the impact of the segment-level aberrations and vibrations in the dark region and provides a stable performance over long exposure times while relaxing the constraints for target contrast. Because it cannot impact angular separations smaller than the segment diffraction limit, it is suitable only for low segment-count mirror telescopes such as the GMT and not for the ELT and TMT. Further applications to the telescopes in the case of island-effect errors will be the subject of a second paper, while this paper focuses on primary-mirror errors caused by segmentation alone.

This paper describes the principle of the RAP design and presents its validation on a simple segmented aperture made of seven circular segments, that is, similar to the GMT configuration, for two coronagraph types: a combination of amplitude RAP and classical APLC, and pure redundant phase RAPs without a second stage. The robustness of these designs is measured for piston-like phasing errors, tip-tilt phasing errors, and missing segments. 

%%%%%%%%%%%%%%%%%%%%%%%%%%%%%%%%%%%%%%%%%%%%%%%%%%%%%%%%
\section{Principle of the RAP}
\label{s:Principle of the redundant apodized pupil}
%%%%%%%%%%%%%%%%%%%%%%%%%%%%%%%%%%%%%%%%%%%%%%%%%%%%%%%%

\subsection{ PASTIS model}
\label{s:Reminders on the PASTIS model}

\begin{figure}%[h]%*} %[h]
   \begin{center}
   \begin{tabular}{c}
   \includegraphics[width=8.5cm]{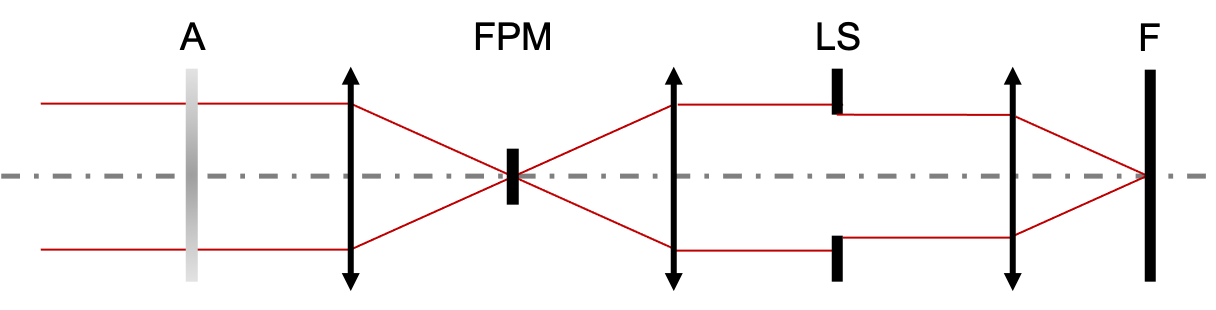}
   \end{tabular}
   \end{center}
   \caption[Corono] 
   { \label{fig:Corono} 
Scheme of a typical coronagraph, composed here of an apodizer (A), an FPM, and a Lyot stop (LS). F stands for for the detector plane. 
}
   \end{figure}

The coronagraphic instrument as described in Fig.\ref{fig:Corono} can be composed of up to three masks: a focal-plane mask (FPM), an apodizer in a pupil plane ahead of the FPM, and a Lyot stop behind the FPM. The PASTIS model \citep{Leboulleux2017, Leboulleux2018}, described in this section, enables deriving the intensity and contrast in the dark hole after a coronagraphic system as a function of the segment-phase errors, with a consequent gain of time in numerical simulation computations compared to end-to-end simulators. It can also been inverted to obtain phasing, polishing, and stability constraints as a function of the target contrast. This is not be presented here but can be found in \cite{Laginja2021}.

%-------------
%-------------

PASTIS is freely available on GitHub \citep{pastis, Laginja2021} and has been extended to combinations of errors and vibrations \citep{Leboulleux2018a}. It has also been applied to LUVOIR \citep{Laginja2019}, and is currently being experimentally validated on the HiCAT testbed at the Space telescope Science Institute \citep{Laginja2020}.

\subsubsection{Hypotheses}
\label{s:Hypotheses}

We consider a segmented aperture space telescope whose primary mirror is composed of hexagonal segments and that is combined with a coronagraph enabling the direct imaging of exoplanets. This coronagraph is responsible for a dark zone in which the possible exoplanet is searched for. PASTIS also requires the following optical conditions: amplitude aberrations and phase aberrations after the focal plane mask of the coronagraph are neglected, and the primary mirror phase aberrations are quite small.

\subsubsection{Expression of the intensity in the dark zone}
\label{s:Expression of the intensity in the dark zone}

When one single Zernike polynomial is applied on the primary mirror segments (such as segment-level piston aberrations for phasing errors), the PASTIS model stipulates that the intensity in the dark zone $I$ at a wavelength $\lambda$ can be expressed as follows:
\begin{equation}
\label{eq:PASTIS}
    I(\mathbf{u}) = \left \Vert \widehat{Z}(\mathbf{u}) \right \Vert ^2 \times \sum_{k_1=1}^{n_{seg}} \sum_{k_2=1}^{n_{seg}} c_{k_1,l} a_{k_1,l} c_{k_2,l} a_{k_2,l} \cos((\mathbf{r_{k_2}} - \mathbf{r_{k_1}} ). \mathbf{u})
,\end{equation}
where $n_{seg}$ is the number of segments in the primary mirror, $\mathbf{u}$ is the position vector in the focal plane, $\widehat{Z}$ is the Fourier transform of the Zernike polynomial $Z$, defined on a hexagonal support, $(c_k)_{k\in \lbrack 1,n_{seg} \rbrack}$ are calibration coefficients taking into account the coronagraph, $(a_k)_{k\in \lbrack 1,n_{seg} \rbrack}$ are the local Zernike coefficients on the segments, and $(\mathbf{r_k})_{k\in \lbrack 1,n_{seg} \rbrack}$ correspond to the position vectors from the center of the pupil to the centers of the segments.

This expression is equivalent to a sum of interference fringes between all pairs of segments modulated by the low-frequency envelope $\left \Vert \widehat{Z}(\mathbf{u}) \right \Vert ^2$, which only depends on the segment shape and on the considered Zernike polynomial. This envelope was studied in \cite{Yaitskova2002} and \cite{Yaitskova2003}, where it was called halo.

\subsection{Theory of segment apodization}
\label{s:Theory of segment apodization}

From Equation \ref{eq:PASTIS}, the overall performance of the coronagraphic system relies on the envelope element $\left \Vert \widehat{Z}(\mathbf{u}) \right \Vert ^2$, which depends on a Zernike polynomial defined from the segment shape support.

We consider piston-like phasing errors below, in which the Zernike polynomial is constant over the segment support. We also call $S$ the apodized segment map ($S$ corresponds to a disk for the GMT and to a hexagon for most other segmented telescopes, e.g., the James Webb Space Telescope or JWST, LUVOIR, the Keck telescope, or the ELT). We then obtain that the envelope $\left \Vert \widehat{Z}(\mathbf{u}) \right \Vert ^2$ is equal to the point spread function (PSF) of one segment. 

From this point, we propose apodizing the segment to modify its PSF and reduce this envelope within the coronagraph dark zone. This apodization reduces the impact of phasing errors on the contrast, and as a consequence, allows larger constraints on the segment-phasing tolerancing.

\begin{figure}
   \begin{center}
   \begin{tabular}{c}
   \includegraphics[width=8.5cm]{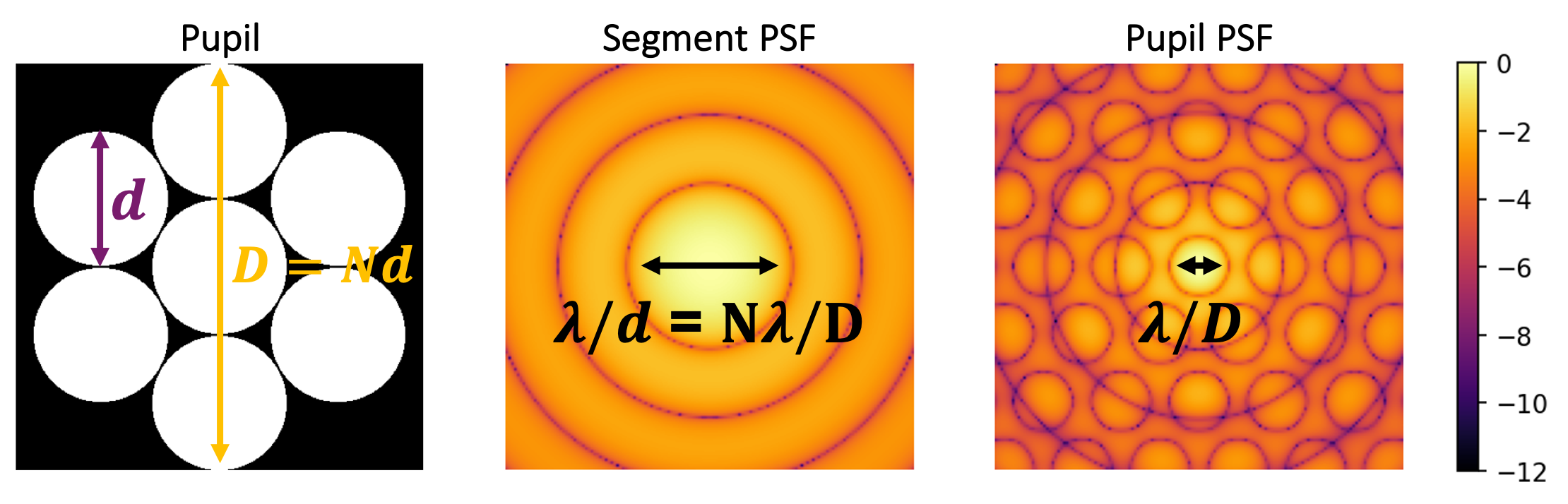}
   \end{tabular}
   \end{center}
   \caption[FigureDefinitions] 
   { \label{fig:FigureDefinitions} 
Impact of the segment on the pupil PSF: (left) Segmented pupil with definitions of $D$ and $d$, where $D=Nd$ (here $N=3$), (center) PSF of one segment on a logarithmic scale, (right) PSF of the overall pupil on a logarithmic scale at the same size as the segment PSF. The relation $D=Nd$ can be observed in the relative sizes of the resolution elements $\lambda/d$ and $\lambda/D$: here $\lambda/d = 3\lambda/D$.}
   \end{figure}

We call $N$ the number of segments along the primary mirror diameter ($N = 3, 5, 7$ for the GMT, JWST, and Keck telescopes), $D$ the diameter of the overall primary mirror, and $d$ the diameter of one segment. Following the relation $D \approx N d$, the envelope $\left \Vert \widehat{Z}(\mathbf{u}) \right \Vert ^2$ is $\sim N$ times larger than the overall primary mirror PSF (see Fig. \ref{fig:FigureDefinitions}). This ratio has to be taken into account when the segment apodization is optimized because the envelope dark zone should cover the coronagraph dark zone.

This ratio also limits the access to small inner working angles: for a coronagraph digging a circular symmetrical dark zone between $\alpha_I \lambda/D$ (inner working angle or IWA) and $\alpha_O \lambda/D$ (outer working angle or OWA), the intensity of the PSF should be minimized between $\sim (\alpha_I/N) \times \lambda/d$ and $\sim (\alpha_O/N) \times \lambda/d$ relative to the central peak intensity.

%-------------
%-------------

%%%%%%%%%%%%%%%%%%%%%%%%%%%%%%%%%%%%%%%%%%%%%%%%%%%%%%%%
\section{Two-step coronagraph design: Validation with APLCs}
\label{s:Numerical validation on a GMT-like aperture}
%%%%%%%%%%%%%%%%%%%%%%%%%%%%%%%%%%%%%%%%%%%%%%%%%%%%%%%%

This section aims to test the proposition of section \ref{s:Theory of segment apodization} in a two-stage design with APLCs by comparing the designs and tolerancing of coronagraphic instruments: one with a classical segmented pupil (CSP), and two with RAP designs.

\subsection{Application cases}
\label{s:Application cases}

The access to small angular separations depends on the number of segments across the pupil diameter $N$, therefore we focus on the GMT pupil, which is composed of seven circular segments, three across the pupil diameter \citep{Fanson2018, Codona2004}. The pupil has a diameter of $24.5$m (equivalent in angular resolution), and each segment is $8.4$m large. 

In the RAP cases, the segments are apodized so that their envelopes reach contrasts of $10^{-4}$ (RAP1 case) and $10^{-5}$ (RAP2 case) within $2.5$ and $5 \lambda/d$, meaning that RAP2 is designed to be $ \text{about ten}$ times more robust than RAP1, which is itself designed to be more robust than the CSP, the pupil made of full circular segments.

As a second step, the three designs (RAP1, RAP2, and CSP) were converted into APLCs able to dig a dark zone down to a contrast below $10^{-7}$ between $7.5$ and $14.5 \lambda/D$. Table \ref{tab:Specs} lists the specifications for this validation case.

\begin{table*}
    \caption{Specifications for the three validation cases.}
    \label{tab:Specs}
    \centering
    \begin{tabular}{l||c|c|c}
        & CSP & RAP1 & RAP2 \\
        \hline \hline
        First step: & / & $2.5-5 \lambda/d$ & $2.5-5 \lambda/d$ \\
        Segment apodization & & $C\sim 10^{-4}$ & $C\sim 10^{-5}$ \\
        \hline
        Second step: & $7.5-14.5 \lambda/D$ & $7.5-14.5 \lambda/D$ & $7.5-14.5 \lambda/D$ \\
        APLC over the pupil & $C\sim 10^{-7}$ & $C\sim 10^{-7}$ & $C\sim 10^{-7}$ \\
    \end{tabular}
\end{table*}

\subsection{Segment and pupil designs}
\label{s:Segment and pupil designs}

To design the RAP1 and RAP2 apodizers, we used the linear relation between the electric fields in the entrance pupil and in the plane of the coronographic image in monochromatic light. For the set maximum contrast in the given dark zone, we searched for the binary apodization, that is, the shaped pupil, that provides the highest transmission, using linear programming methods. We used an AMPL (A Mathematical Programming Language) \citep{Fourer2002} code to define the optimization problem presented in \cite{Carlotti2013} and the Gurobi solver to find the optimal solution \citep{gurobi}.

\begin{figure}%[h]%*} %[h]
   \begin{center}
   \begin{tabular}{c}
   \includegraphics[width=8.5cm]{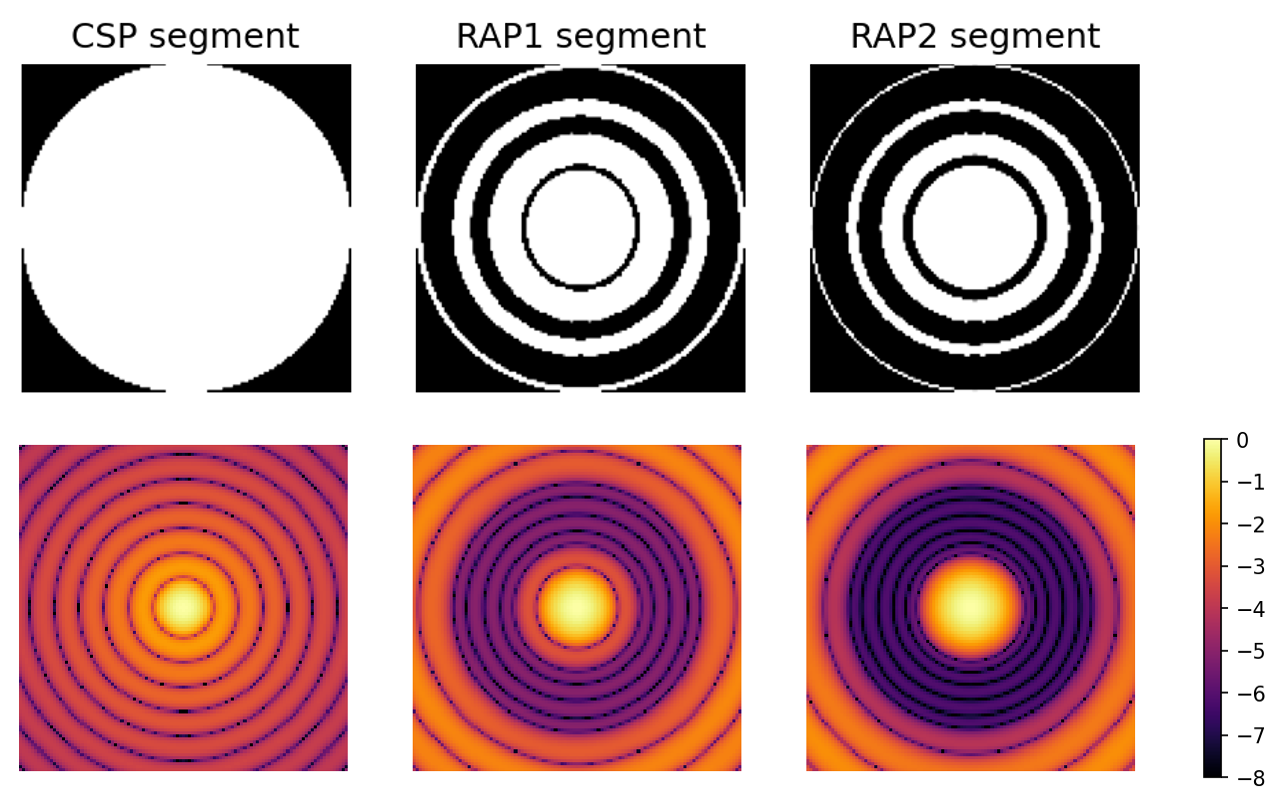}
   \end{tabular}
   \end{center}
   \caption[APLC_SegmentsAndSegmentPSFs] 
   { \label{fig:APLC_SegmentsAndSegmentPSFs} 
(top) Segments and (bottom) associated PSFs in logarithmic scale considered for the numerical validation of the RAP concept: (left) No apodization, (center) segment optimized for an envelope contrast of $10^{-4}$ (RAP1 case), and (right) segment optimized for an envelope contrast of $10^{-5}$ (RAP2 case).}
   \end{figure}

The outcome segment apodizations are shown in Fig.~\ref{fig:APLC_SegmentsAndSegmentPSFs} (center and right) with their PSFs, which also correspond the low-order envelope of the full segmented pupil PSFs.

%-------------
   %*} 
%-------------

\begin{figure}%[h]%*} [h]
   \begin{center}
   \begin{tabular}{c}
   \includegraphics[width=8.5cm]{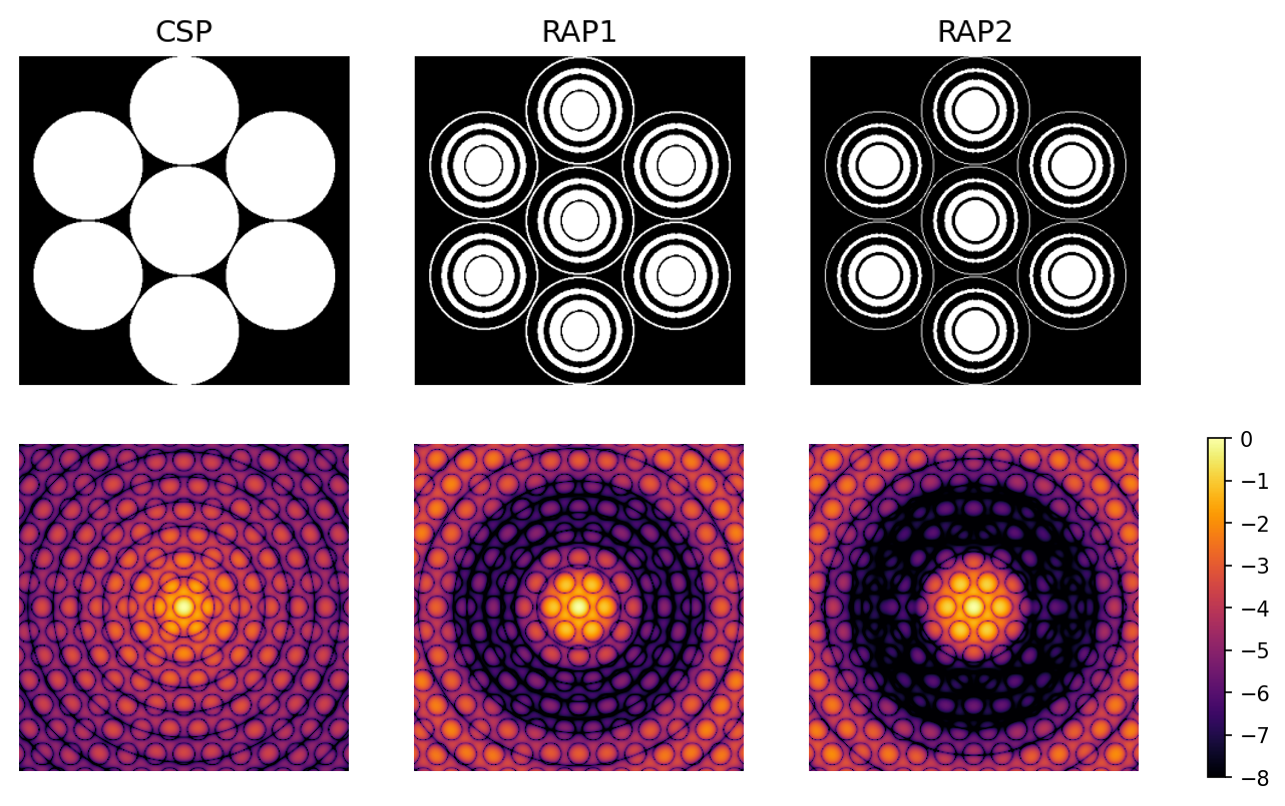}
   \end{tabular}
   \end{center}
   \caption[APLC_PupilsAndPupilPSFs] 
   { \label{fig:APLC_PupilsAndPupilPSFs} 
(top) Pupils and (bottom) associate PSFs in logarithmic scale considered for the numerical validation of the RAP concept: (left) CSP case (reference), (center) RAP optimized for an envelope contrast of $10^{-4}$ (RAP1 case), and (right) RAP optimized for an envelope contrast of $10^{-5}$ (RAP2 case).}
   \end{figure}

The full CSP and RAP pupils are obtained by concatenating these segments within an architecture equivalent to the GMT primary mirror, that is, they consist of seven segments. The final pupil designs and their PSFs are plotted in Fig.~\ref{fig:APLC_PupilsAndPupilPSFs}, and the RAP transmissions compared to the CSP transmission are indicated in Table \ref{tab:Transmissions}. The low-order segment envelopes are also clearly visible in these PSFs.

%-------------
   %*} 
%-------------

\begin{table}[!htb]
    \caption{Comparison of the apodizer transmissions and throughputs of the APLC designs.}
    \label{tab:Transmissions}
    \centering
    \begin{tabular}{l||c|c|c}
        & CSP & RAP1 & RAP2 \\
        \hline \hline
        Transmission after step 1 & $100 \%$ & $50.4 \%$ & $39.2 \%$ \\
        \hline
        Transmission after step 2 & $72.1 \%$ & $47.0 \%$ & $34.9 \%$ \\
        \hline
        Throughput after step 2 & $52.2 \%$ & $22.6 \%$ & $12.7 \%$ \\
    \end{tabular}
\end{table}

\subsection{Coronagraph designs}
\label{s:Coronagraph design}

These redundant pupils were used in a second step as a starting point to design APLCs: Each of these redundant apodizers was set as the input of a second optimization instead of the original GMT-like pupil. This optimization problem was solved by generating the apodizer from the input pupil (or from the redundant apodizer in our case) that maximizes the transmission for a given contrast in a defined dark zone in the coronagraphic image for given FPM and Lyot stop geometries \citep{N'Diaye2015, N'Diaye2016, Zimmerman2016, N'Diaye2018, Por2020}. We also used linear programming methods to determine our solution using a Gurobi solver. The output apodizer patterns then include the redundant apodizer patterns of Fig.\ref{fig:APLC_PupilsAndPupilPSFs}. This output apodizer also consists of the final apodizer of the APLC design as in Fig.\ref{fig:Corono}, which we use below.

The dark zone and the target contrast were set according to the specifications of the second step in Table \ref{tab:Specs}, meaning a target contrast of $10^{-7}$ between $7.5$ and $14.5 \lambda/D$. The CSP and the RAP APLCs include the same FPM of a $6.5 \lambda/D$ radius and the same Lyot stop (circular Lyot stop with a Lyot ratio of $80\%$ ) and were all optimized in polychromatic light over a spectral bandwidth of $10\%$ .

\begin{figure}%[h]%*} [h]
   \begin{center}
   \begin{tabular}{c}
   \includegraphics[width=8.5cm]{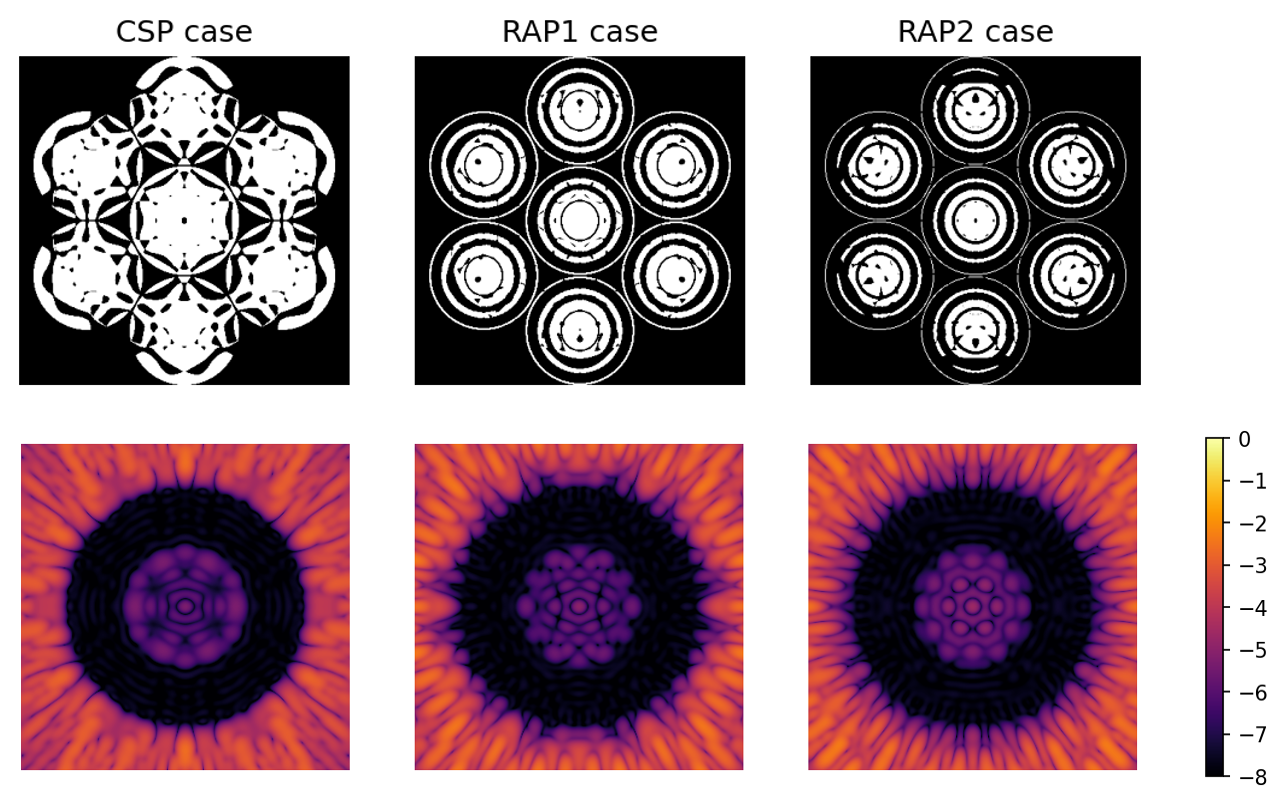}
   \end{tabular}
   \end{center}
   \caption[APLC_ApodsAndApodPSFs] 
   { \label{fig:APLC_ApodsAndApodPSFs} 
(top) Apodizers and (bottom) coronagraphic PSFs in logarithmic scale considered for the numerical validation of the RAP concept: (left) CSP case, and (right) RAP case.}
   \end{figure}

The apodizers and the corresponding coronagraphic PSFs are plotted in Fig.\ref{fig:APLC_ApodsAndApodPSFs}, and the transmissions of the apodizers are indicated in Table \ref{tab:Transmissions} with the planet throughputs as defined in \cite{Por2020}.

%-------------
   %*} 
%-------------

\subsection{Tolerancing and constraint study}
\label{s:Tolerancing and constraint study}

The objective of this section is to test and compare the robustness of these designs to piston-like phasing aberrations, to which the RAPs are designed to be robust.

\begin{figure}%[h]%*} [h]
   \begin{center}
   \begin{tabular}{c}
   \includegraphics[width=8.5cm]{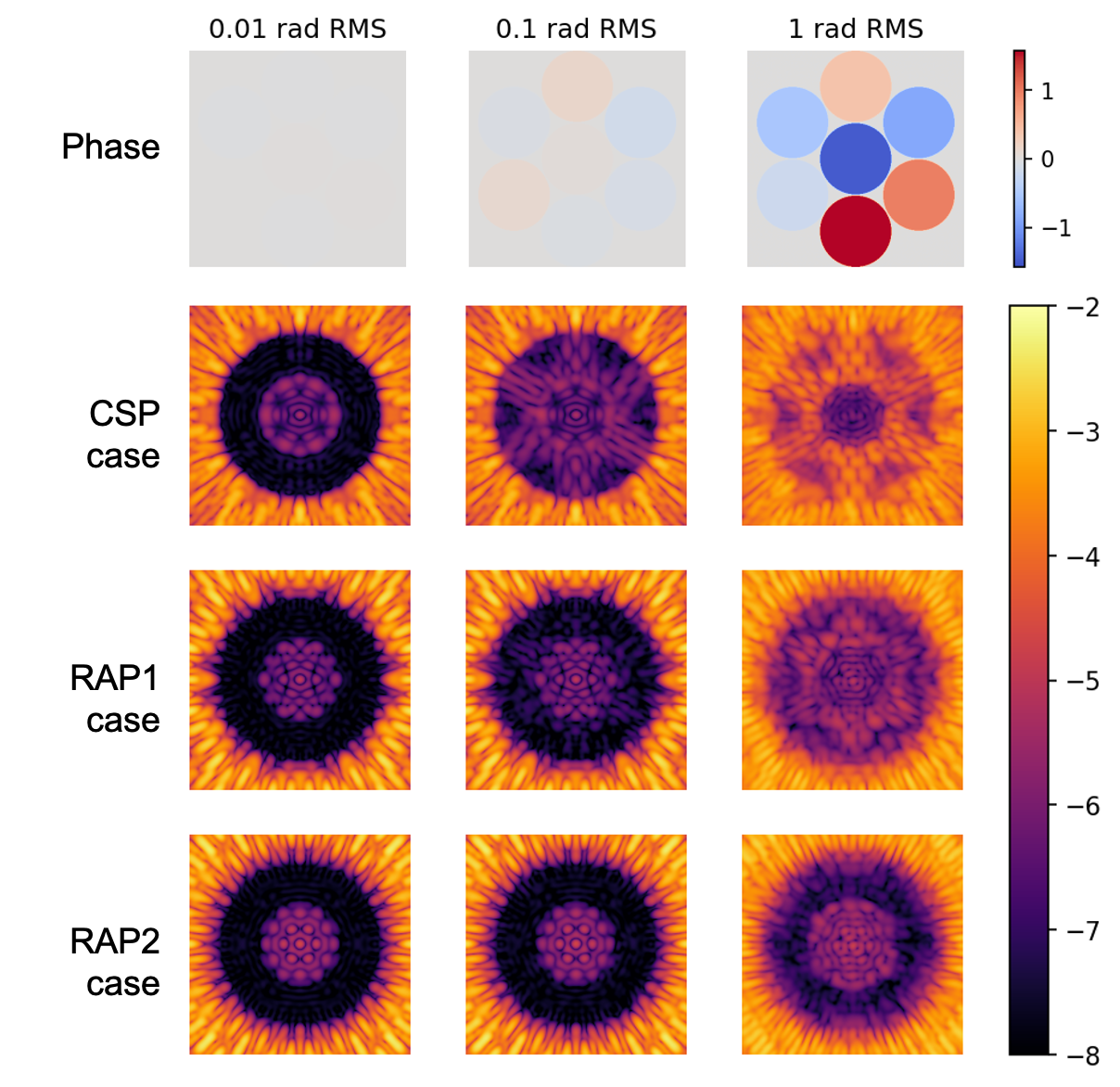}
   \end{tabular}
   \end{center}
   \caption[FigureGMTCase_ErrorBudget] 
   { \label{fig:FigureGMTCase_ErrorBudget} 
Robustness to phasing errors. (top) Phase maps in radians applied on the pupil with three different amplitudes of phasing errors: (left) 0.01 rad RMS, (center) 0.1 rad RMS, and (right) 1 rad RMS. (bottom) PSFs in logarithmic scale with the three designs and for the three phase maps above.}
   \end{figure}

Fig.\ref{fig:FigureGMTCase_ErrorBudget} compares the PSFs when different levels of phasing aberrations are applied to the pupil. The PSFs seem less impacted by aberrations behind the RAP instruments than behind the CSP instrument.

%-------------
   %*} 
%-------------

\begin{figure}
   \begin{center}
   \begin{tabular}{c}
   \includegraphics[width=8.5cm]{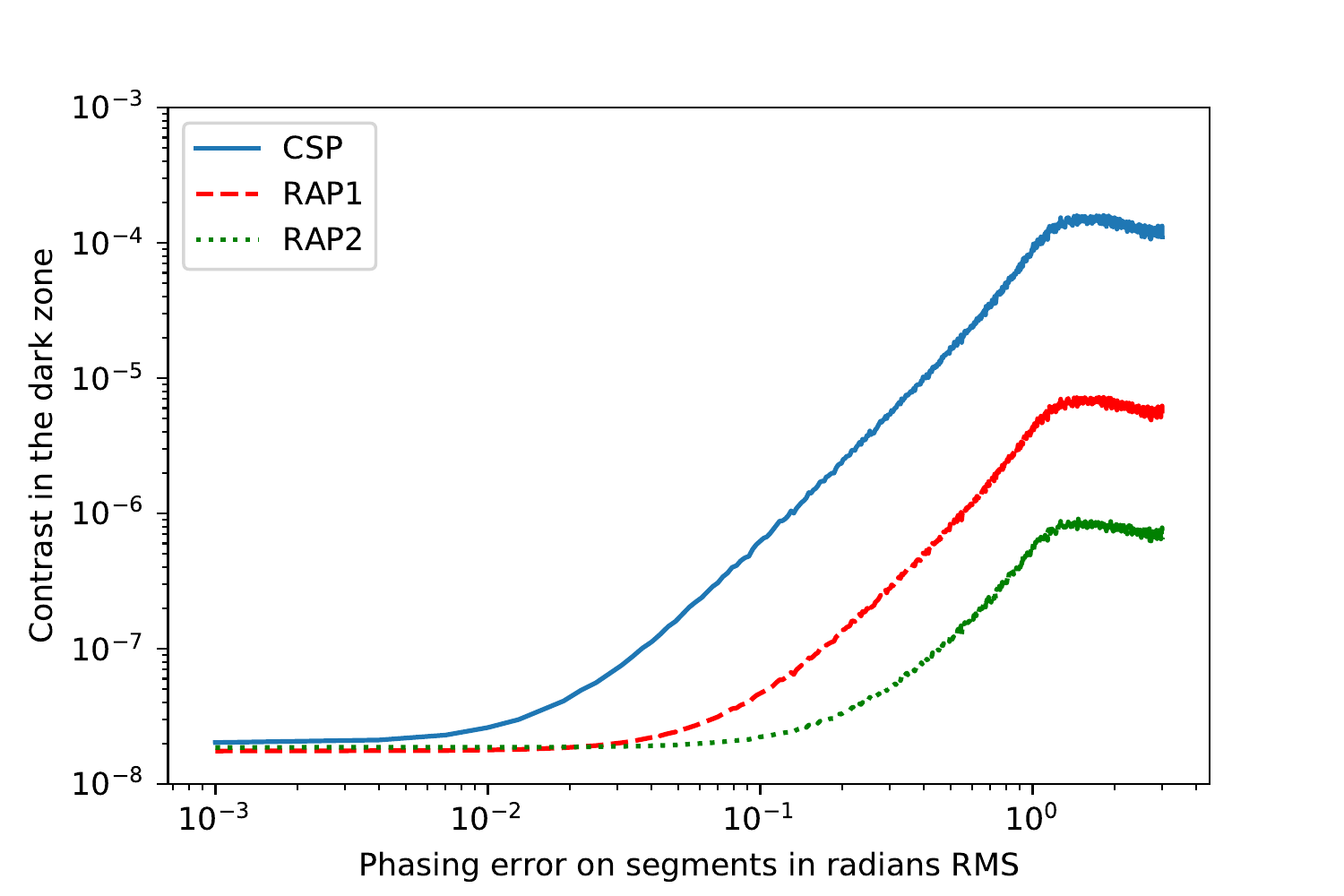}
   \end{tabular}
   \end{center}
   \caption[FigureGMTCase_HockeyCross] 
   { \label{fig:FigureGMTCase_HockeyCross} 
Evolution of the contrast with the phasing-error amplitude: (blue) for the CSP design, (red) for the RAP1 design, and (green) for the RAP2 design. 1000 different error amplitudes are considered between 1 mrad RMS and 3 rad RMS. For each of them, 100 phasing errors are simulated and propagated, 100 PSFs normalized by their aberrated direct-image peak are computed, and the average of the 100 resulting dark hole contrasts is plotted.}
   \end{figure}

More generally, we computed the average contrast in the coronagraph dark region for a wide range of phasing aberrations, testing 1000 phasing-error amplitudes between 0.001 and 3 radians RMS and 100 random phasing-errors per error amplitude. The curves are plotted in Fig.\ref{fig:FigureGMTCase_HockeyCross} and indicate an increasing robustness for the CSP case, the RAP1 case, and the RAP2 case. The average contrast is $20$ ($137$) times less deteriorated with the RAP1 (RAP2) design than with the CSP design for a given phasing-error amplitude (computed at $0.5$ rad RMS).

In addition, the constraints in terms of phasing were released by a factor of $4.7$ for RAP1 and by a factor of $13$ for RAP2: a $10^{-7}$ contrast imposes a constraint of $37$, $170$, and $463$ mrad RMS of phasing errors with the CSP, the RAP1, and the RAP2 designs.

Fig.\ref{fig:FigureGMTCase_HockeyCross} also indicates that the three coronagraphic instruments are not equally robust to high-amplitude aberrations because they enable reaching different contrast plateaus: $1.1 \times 10^{-4}$, $5.2 \times 10^{-6}$, and $6.4 \times 10^{-7}$ for CSP, RAP1, and RAP2, respectively. Before these plateaus, the three curves reach a local maximum: this occurs when the segment-level pistons reach almost $\pm \pi$, that is, right before the phases start to be optically shifted by $2 \pi$. When the phasing error RMS continues to increase above this local maximum, the segment-level optical ($2 \pi$-shifted) pistons quickly become fully random in the range $[ - \pi, \pi]$, which explains the plateaus.

%-------------

\subsection{Robustness to other errors}
\label{s:Comments on robustness to other errors}

In this section, we study the robustness of the CSP and RAP designs to other errors caused by segmentation in phase and amplitude: tip-tilt phasing errors, and missing segments in the pupil.

\begin{figure}%[h]%*} [h]
   \begin{center}
   \begin{tabular}{c}
   \includegraphics[width=8.5cm]{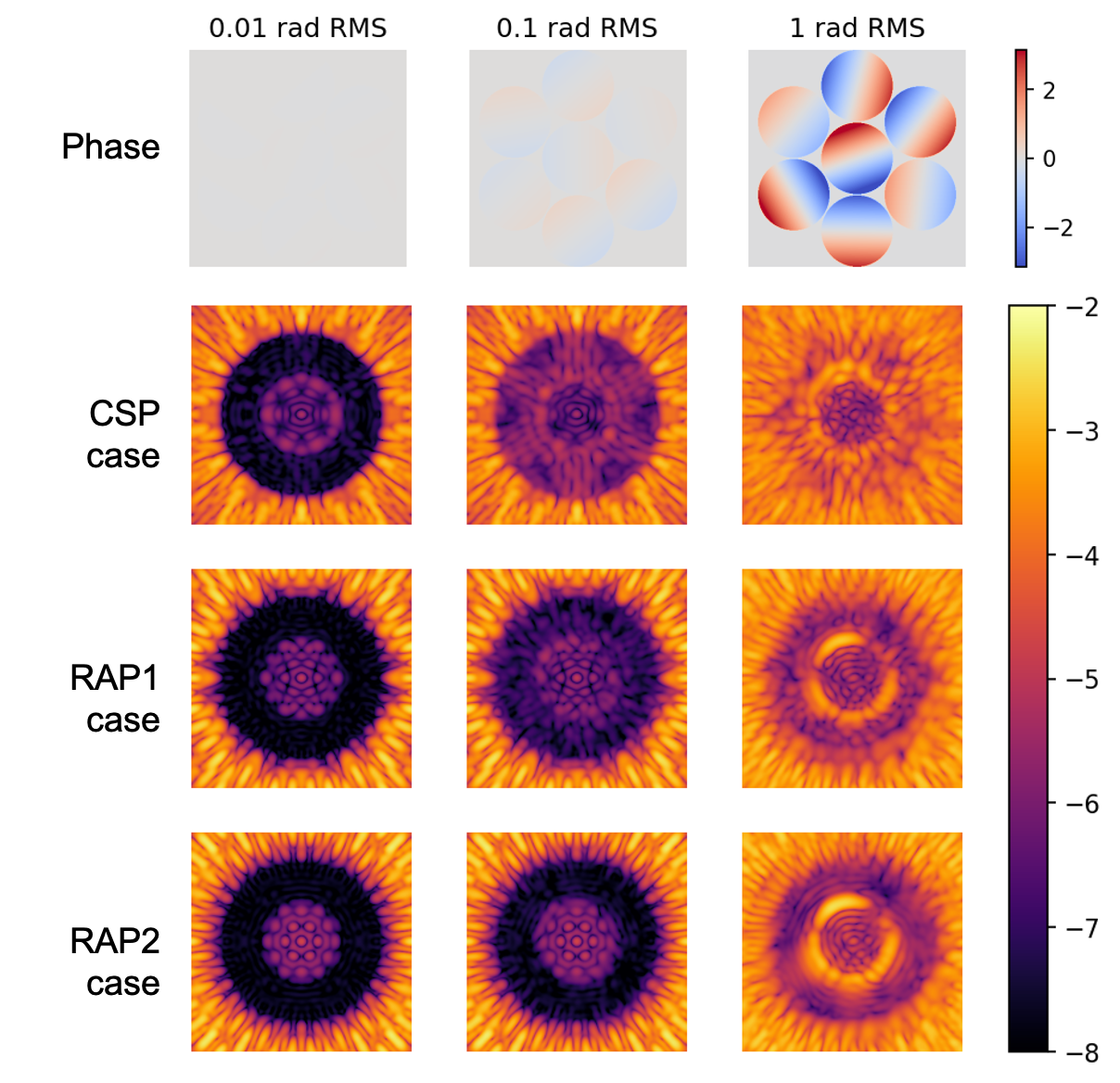}
   \end{tabular}
   \end{center}
   \caption[APLCCase_TTErrorBudget] 
   { \label{fig:APLCCase_TTErrorBudget} 
Robustness to tip-tilt phasing errors. (top) Phase maps in radians applied on the pupil with three different amplitudes of tip-tilt phasing errors: (left) 0.01 rad RMS, (center) 0.1 rad RMS, and (right) 1 rad RMS. (bottom) PSFs in logarithmic scale with the three designs and for the three phase maps above.}
   \end{figure}

As a first test, segment-level tip and tilt errors were added on the primary mirrors. Fig.\ref{fig:APLCCase_TTErrorBudget} shows the PSFs for three levels of tip-tilt phasing errors ($0.01$, $0.1$, and $1$ rad RMS): the RAP designs appear to have a lower impact on the coronagraphic PSF than the CSP design. There is no impact at $0.01$ for the two RAP designs and an impact of up to $0.1$ rad RMS for RAP2. 

\begin{figure}%[h]%*} [h]
   \begin{center}
   \begin{tabular}{c}
   \includegraphics[width=8.5cm]{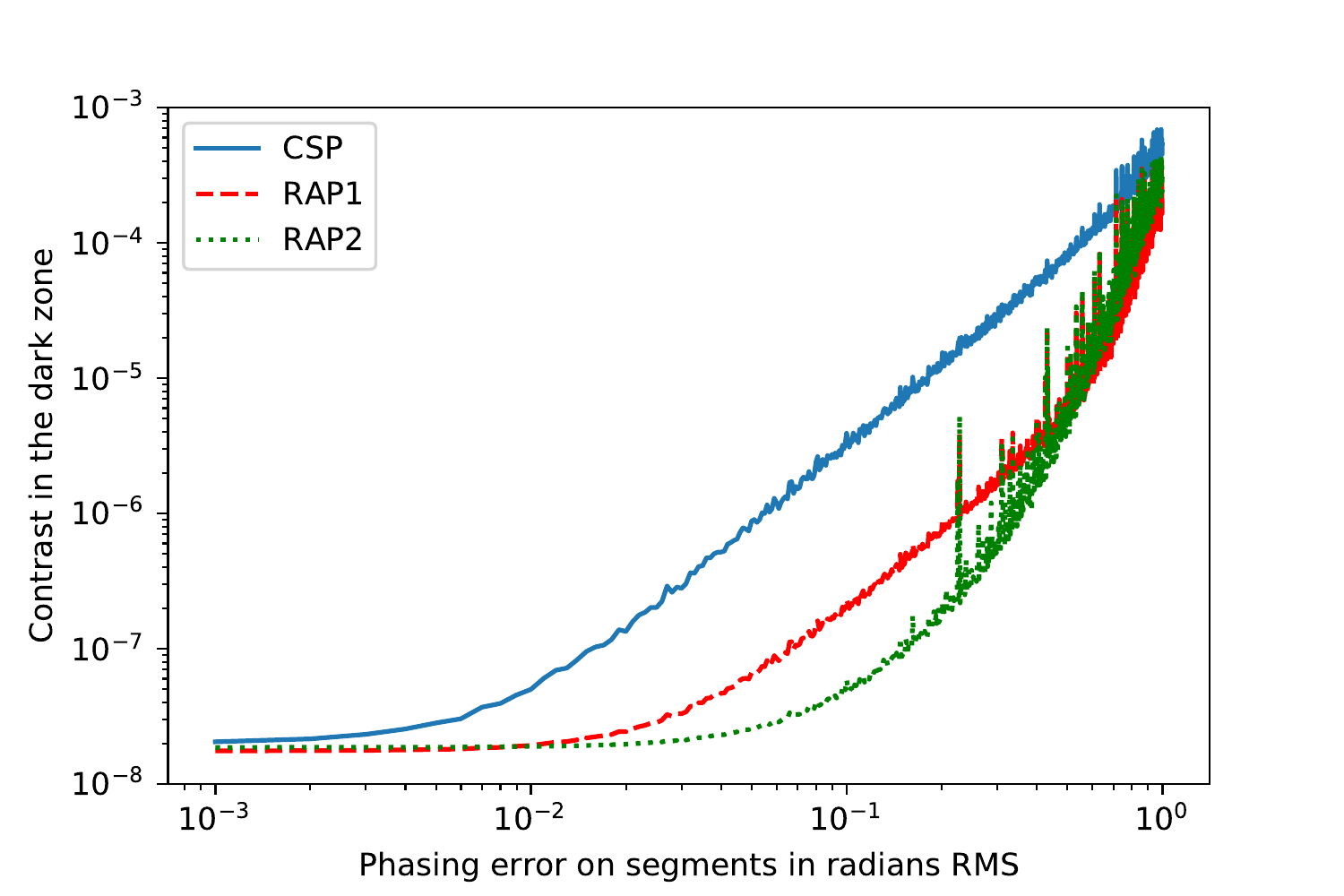}
   \end{tabular}
   \end{center}
   \caption[APLC_TTHockeyCross] 
   { \label{fig:APLC_TTHockeyCross} 
Evolution of the contrast with the tip-tilt phasing error amplitude: (blue) For the CSP design, (red) for the RAP1 design, and (green) for the RAP2 design. 1000 different error amplitudes are considered between 1 mrad RMS and 1 rad RMS. For each of them, 100 phasing errors are simulated and propagated, 100 PSFs normalized by their aberrated direct-image peak are computed, and the average of the 100 resulting dark hole contrasts is plotted.}
   \end{figure}

Once again, this verification was extended to a wide range of segment-level tip-tilt aberration amplitudes. This is plotted in Fig.\ref{fig:APLC_TTHockeyCross}. We employed 1000 different error amplitudes and 100 random phase maps per amplitude. With the RAP1 design, the impact of the errors on the contrast in the coronagraph dark zone remains at its optimal level until $\sim 20$ mrad RMS before it increases ($\sim 50$ mrad RMS for the RAP2 design), and it stays below the CSP design in the whole aberration range considered here. Accessing a mean contrast in the dark zone of $10^{-7}$, for instance, imposes tip-tilt phasing constraints down to $16$ mrad RMS with the CSP design, but $68$ and $157$ mrad RMS with the RAP1 and RAP2 designs. This is a difference of a factor $4$ and $10$. Because the tip and tilt envelopes are different from the piston envelope \citep{Leboulleux2018} and because the segment apodization was optimized on the piston envelope alone, the difference in robustness with piston-like errors is expected, but the improvement is still significant.

%-------------
%-------------

%-------------
   %*} 
%-------------

\begin{figure}%[h]%*} [h]
   \begin{center}
   \begin{tabular}{c}
   \includegraphics[width=8.5cm]{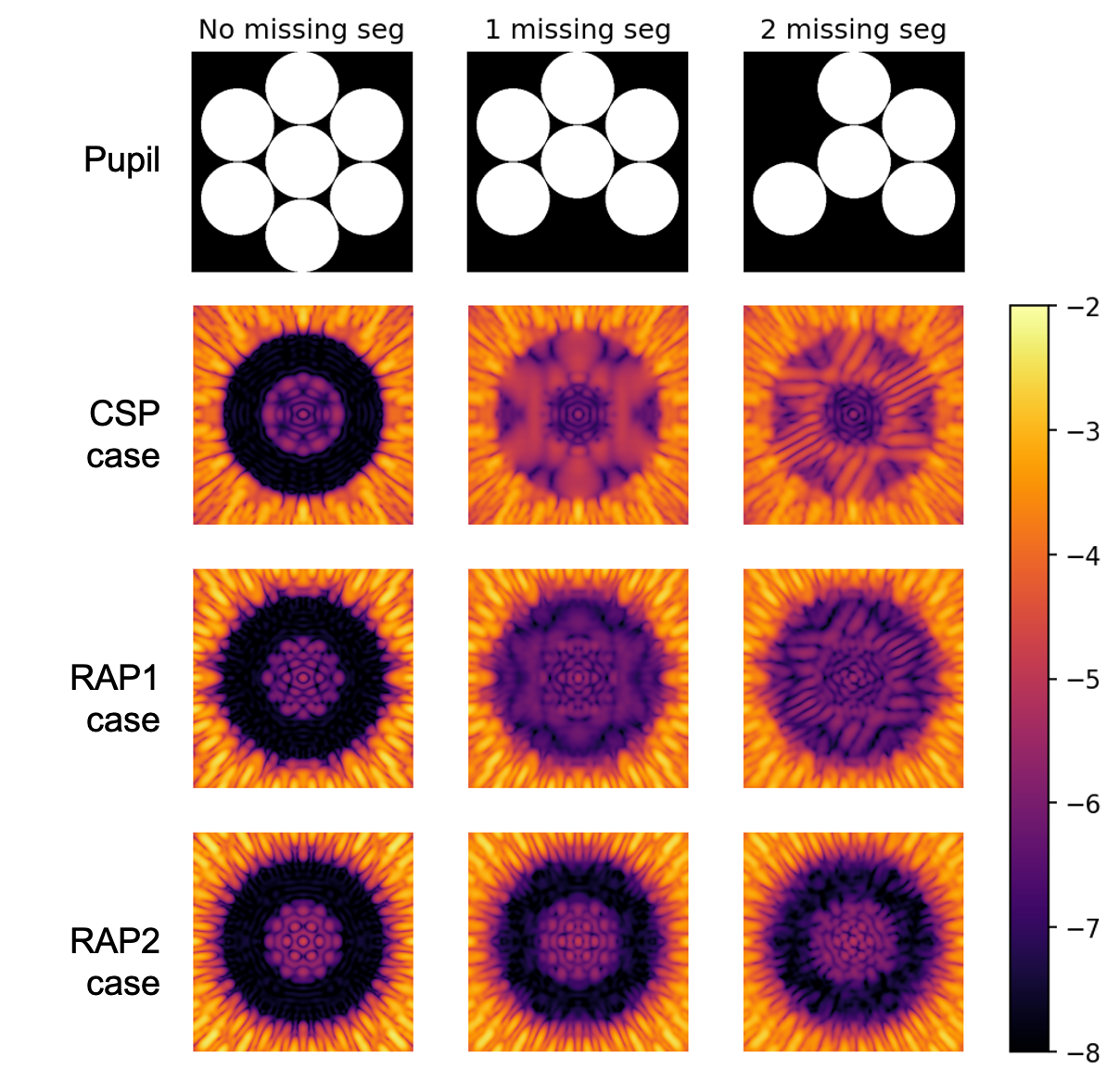}
   \end{tabular}
   \end{center}
   \caption[APLC_MissingSegmentsErrorBudget] 
   { \label{fig:APLC_MissingSegmentsErrorBudget} 
Impact of missing segments on the CSP and RAP design PSFs and their dark-zone contrasts: (top) Pupil configuration without (left) missing segment, with (center) one missing segment, and with (right) two missing segments. (bottom) PSFs for the CSP, RAP1, and RAP2 cases when subject to the pupil configuration above.}
   \end{figure}

As a second test, we removed one and two segments from the primary mirror and studied the impact of these missing segments on the coronagraphic PSFs in the CSP and the RAP configurations. Fig.\ref{fig:APLC_MissingSegmentsErrorBudget} and Table \ref{tab:APLC_MissingSegments} illustrate this effect: the contrast is degraded by a factor of $1160$, $70$, and $8.3$ for CSP, RAP1, and RAP2 when no segment and when two segments are missing, and the RAP designs are more robust to missing segments and enable maintaining good contrast, in contrast with the CSP design. This is explained by the fact that in the RAP cases, the low-order envelope is not impacted by missing segments, but only by missing baselines in the pupil, in contrast with the CSP case.

%-------------
%-------------

\begin{table}%[!htb]
    \caption{Contrasts without and with one and two missing segments in the pupil.}
    \label{tab:APLC_MissingSegments}
    \centering
    \begin{tabular}{l||c|c|c}
        & 0 mis. seg. & 1 mis. seg. & 2 mis. seg. \\
        \hline \hline
        CSP & $2.0 \times 10^{-8}$ & $7.6 \times 10^{-6}$ & $2.3 \times 10^{-5}$ \\
        \hline
        RAP1 & $1.8 \times 10^{-8}$ & $4.5 \times 10^{-7}$ & $1.2 \times 10^{-6}$ \\
        \hline
        RAP2 & $1.9 \times 10^{-8}$ & $6.7 \times 10^{-8}$ & $1.6 \times 10^{-7}$ \\
    \end{tabular}
\end{table}

\subsection{Validation conclusions}
\label{s:Validation conclusions}

The objective of this section was to design redundant apodizers combined with APLCs and to compare their robustness to phasing errors with a classical design.

In this validation case, the segment was apodized to deepen its PSF envelope in the dark zone ($2.5-5 \lambda/d$) down to $10^{-4}$ and $10^{-5}$ to study two levels of robustness. This leads to a gain of factors of $4.7$ and $13$ for the phasing constraints for a given contrast in the coronagraph dark zone ($7.5-14.5 \lambda/D$).

This performance is slightly impacted by the second-step apodization, which affects the segment envelope by slightly modifying the segment apodization. In addition, the RAP is also optimized on a monochromatic light, while its robustness is checked on a $10 \%$ polychromatic light, which can also cause a slight loss of robustness.

Overall, the main drawback of the RAP designs is its loss in transmission and throughput compared to the CSP design. It also limits the access to small IWAs because shaped pupils can hardly access separations lower than $\sim 1.8-2 \lambda/d$ (i.e., $\sim 6 \lambda/D$) \citep{Carlotti2012}. A solution is using APP designs instead of shaped pupils: they can access smaller angular separations with a higher throughput \citep{Por2017}.

%%%%%%%%%%%%%%%%%%%%%%%%%%%%%%%%%%%%%%%%%%%%%%%%%%%%%%%%
\section{One-step coronagraph design: Application to APPCs}
\label{s:Application to APP coronagraphs}
%%%%%%%%%%%%%%%%%%%%%%%%%%%%%%%%%%%%%%%%%%%%%%%%%%%%%%%%

We now focus on RAPs built from the first step only that combine coherently on the detector plane without further design at the full pupil scale. In Fig.\ref{fig:Corono} this corresponds to a coronagraph composed of the apodizer alone, without a focal-plane mask and Lyot stop. In order to reduce the IWA, they are based on redundant APPs \citep{Codona2007, Kenworthy2010, Por2017} instead of shaped pupils. They also aim for more modest contrasts, and no second step is used to heighten the contrast.

\subsection{Segment and pupil designs}
\label{s:Segment and pupil design2}

The segment phase was optimized to reach an envelope contrast of $10^{-5.5}$ at most between $2$ and $4.14 \lambda/d$ in a one-sided dark zone on one hand, and in a two-sided dark zone on the other hand. We used the same AMPL code as earlier for SPs to obtain APPs in a circular segment, with the following constraints \citep{Carlotti2013}:

\begin{equation}
    \forall \vec{u}, A_I(\vec{u}) = \begin{cases} -A_R(-\vec{u}) \mbox{ for a one-sided dark zone} \\
                                                  -A_R(\vec{u}) \mbox{ for a two-sided dark zone,} \\
                                    \end{cases}
\end{equation}
where $A$ is the segment apodization, and $A_R$ and $A_I$ are its real and imaginary parts.

\begin{figure}%[h]%*} [h]
   \begin{center}
   \begin{tabular}{c}
   \includegraphics[width=8.5cm]{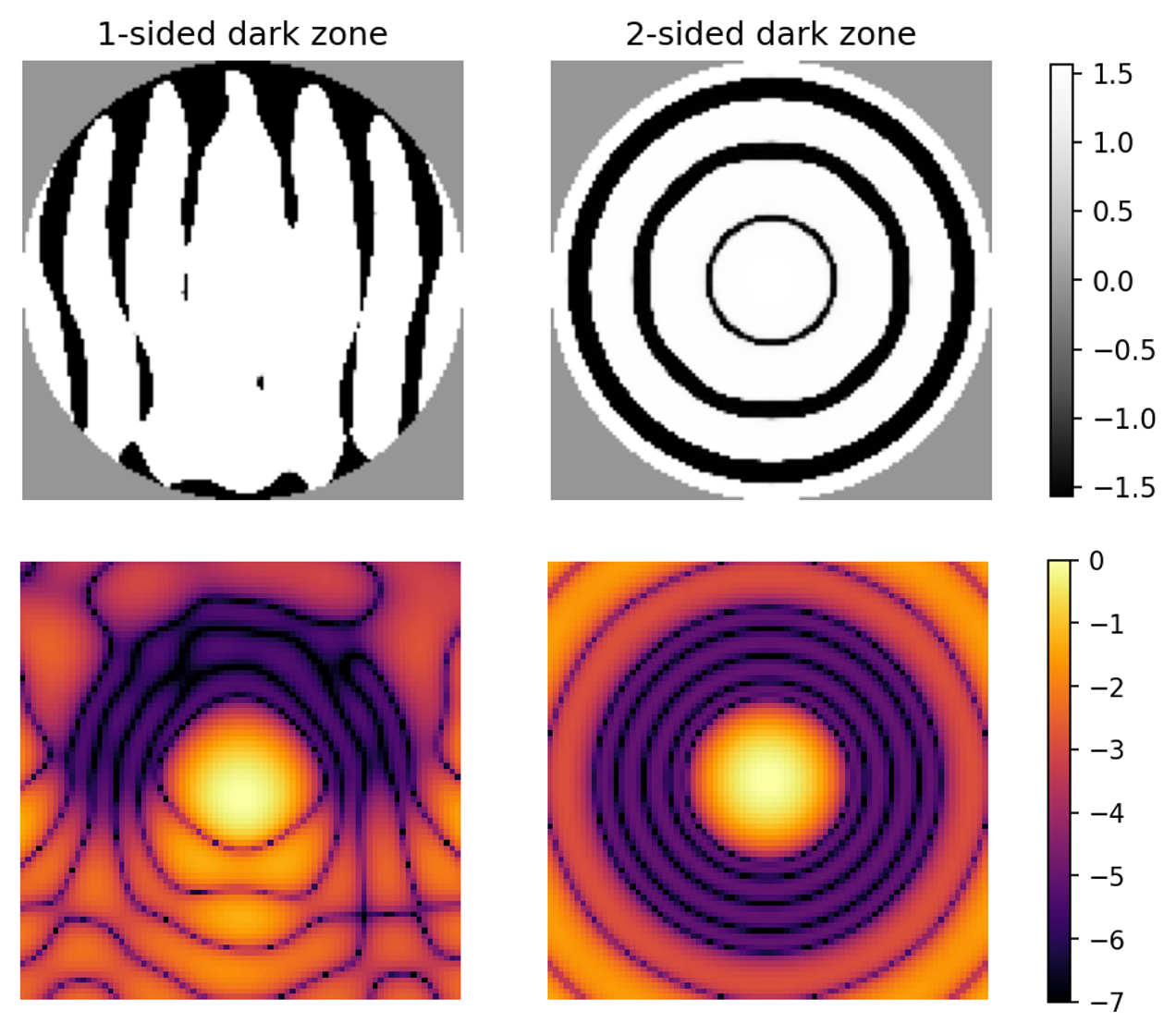}
   \end{tabular}
   \end{center}
   \caption[APP_SegmentAndSegmentPSF] 
   { \label{fig:APP_SegmentAndSegmentPSF} 
Segment phase apodizations: (top) Segment phases in radians, optimized for an envelope contrast of $10^{-5.5}$ in the dark zones and (bottom) associated PSFs in logarithmic scale, for (left) a one-sided dark zone and (right) a two-sided dark zone.}
   \end{figure}

\begin{figure}%[h]%*} [h]
   \begin{center}
   \begin{tabular}{c}
   \includegraphics[width=8.5cm]{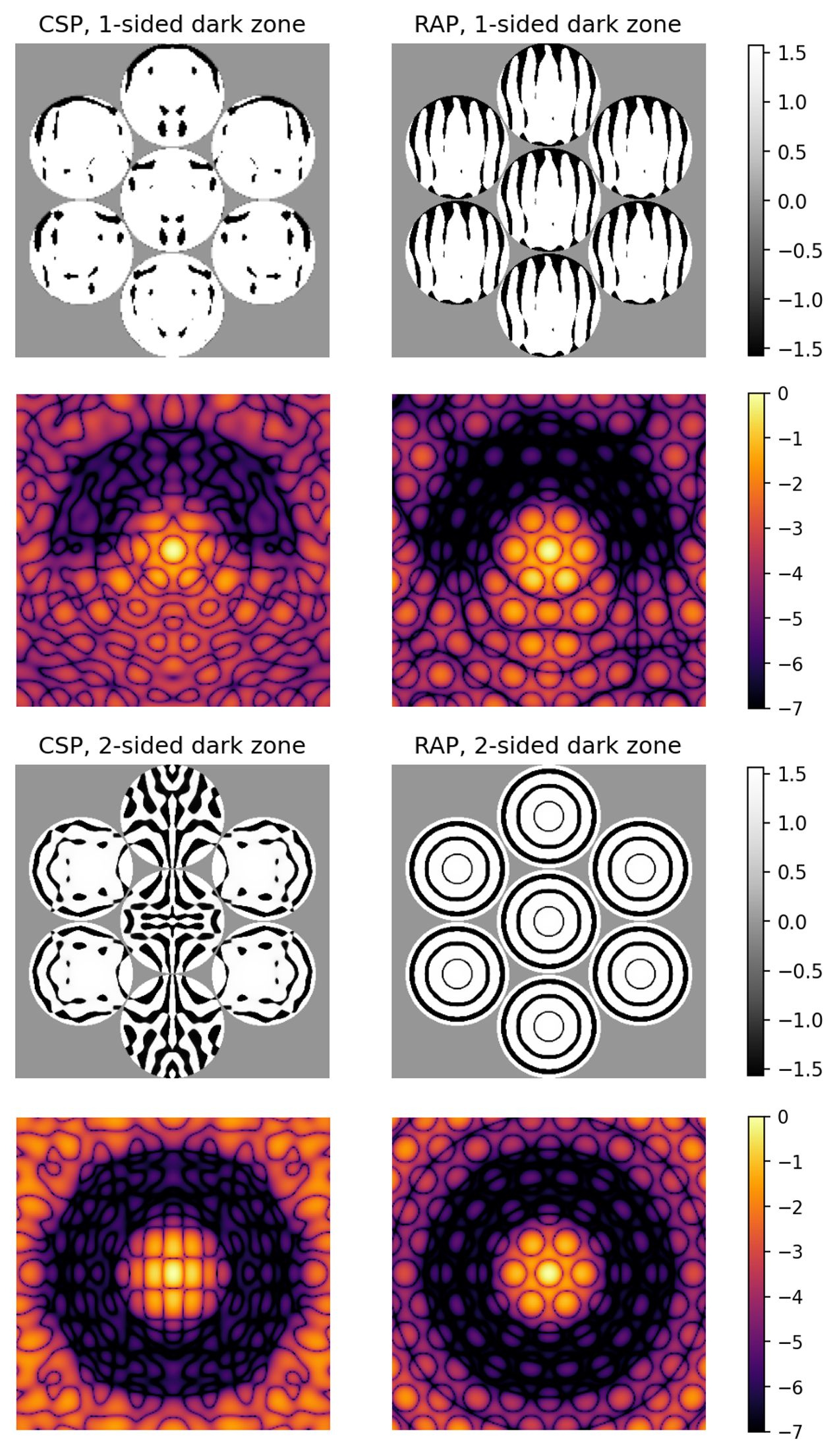}
   \end{tabular}
   \end{center}
   \caption[APP_PupilsAndPupilPSFs] 
   { \label{fig:APP_PupilsAndPupilPSFs}
(line 1) APPs in radians, and (line 2) associated PSFs in logarithmic scale for a one-sided dark zone, (line 3) APPs in radians, and (line 4) associated PSFs in logarithmic scale for a two-sided dark zone: (left) Reference cases with an APP optimized for the whole CSP pupil and for (right) RAP designs.}
   \end{figure}

In Fig.\ref{fig:APP_SegmentAndSegmentPSF} the outcome apodization phases (top) and their associated PSFs (bottom) are shown. The associated RAPs, called RAP1 (one-sided dark zone) and RAP2 (two-sided dark zone), are shown in Fig.\ref{fig:APP_PupilsAndPupilPSFs} with their associated PSFs (right). These PSFs are still optimized to reach a contrast of $10^{-5.5}$ in one-sided and two-sided dark zones between $5.5$ and $12 \lambda/D$.  RAP1 is not fully left-right symmetrical: this is first due to the fact that the Gurobi optimization was made for the entire segment without a symmetry constraint, in contrast with the apodizers for two-sided dark zones that were only computed for a quarter and were then extended symmetrically to the rest of the segment or pupil, and second to the sampling of the segment that was set low to keep the full pupil size small enough for simulations.

%-------------
%-------------

%-------------
%-------------

These two RAPs are compared below with APPs that were classically obtained on the entire GMT-like pupil, also optimized for the same specifications. This means a contrast of $10^{-5.5}$ between $5.5$ and $12 \lambda/D$ in the one-sided dark zone and the two-sided dark zone. These CSP APPs, called CSP1 and CSP2, are shown in Fig.\ref{fig:APP_PupilsAndPupilPSFs} with their associated PSF (left). Table~\ref{tab:APP} also lists the planet throughputs of the four APPs considered in this section. For a one-sided dark zone, the throughput is deteriorated by the redundance of the apodization, which is mainly due to segment-level tips that slightly shift the envelope. For a two-sided dark zone in contrast, the throughput is better with the RAP design than with the CSP.

\begin{table}%[!htb]
    \caption{Planet throughputs of the four APP cases.}
    \label{tab:APP}
    \centering
    \begin{tabular}{c|c||c|c}
        CSP1 & RAP1 & CSP2 & RAP2 \\
        \hline \hline
        $64.1 \%$ & $26.7 \%$ & $13.7 \%$ & $17.3 \%$ \\
    \end{tabular}
\end{table}

\subsection{Tolerancing and constraint study}
\label{s:Tolerancing and constraint study2}

\begin{figure}%[h]%*} [h]
   \begin{center}
   \begin{tabular}{c}
   \includegraphics[width=8.5cm]{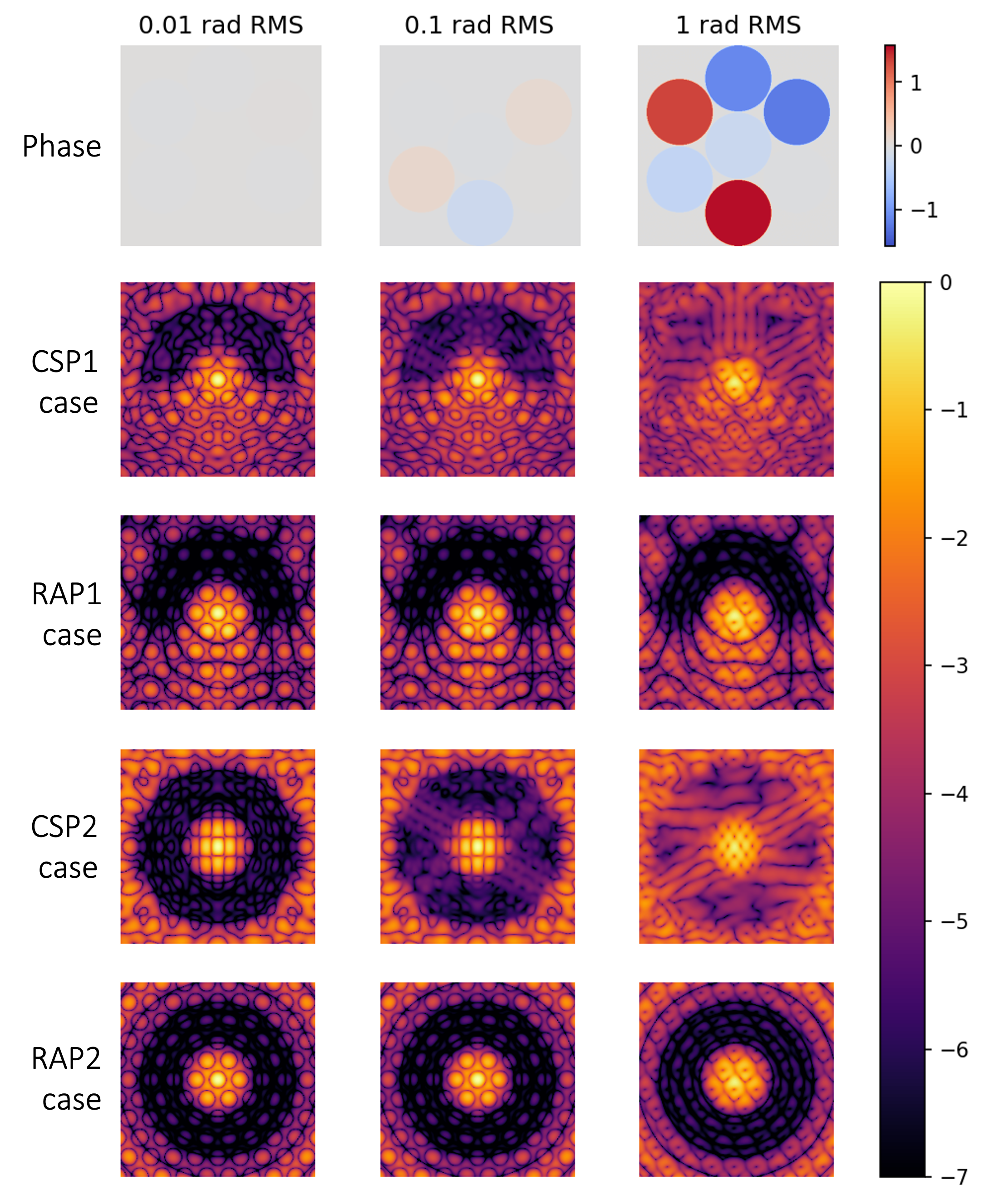}
   \end{tabular}
   \end{center}
   \caption[APP_ErrorBudget] 
   { \label{fig:APP_ErrorBudget} 
Robustness to phasing errors: (top) Phasing errors applied on the GMT-like pupil for three different amplitudes: 0.01 rad RMS, 0.1 rad RMS, and 1 rad RMS. (bottom) PSFs in logarithmic scale with the four APP designs CSP1, RAP1, CSP2, and RAP2 with the aberrations above.}
   \end{figure}

A first comparison of the CSP and RAP design robustness to piston-like phasing errors is shown in Fig.\ref{fig:APP_ErrorBudget}: Three different phasing aberrations were simulated from 0.01 to 1 rad RMS of error (top), with the associated PSFs through the four different APPs (bottom). The average contrasts in the dark zones of the PSFs issued from the RAP designs are almost not impacted by phasing aberrations, in contrast with those from the CSP designs.

%-------------
%-------------

\begin{figure}%[h]%*} [h]
   \begin{center}
   \begin{tabular}{c}
   \includegraphics[width=8.5cm]{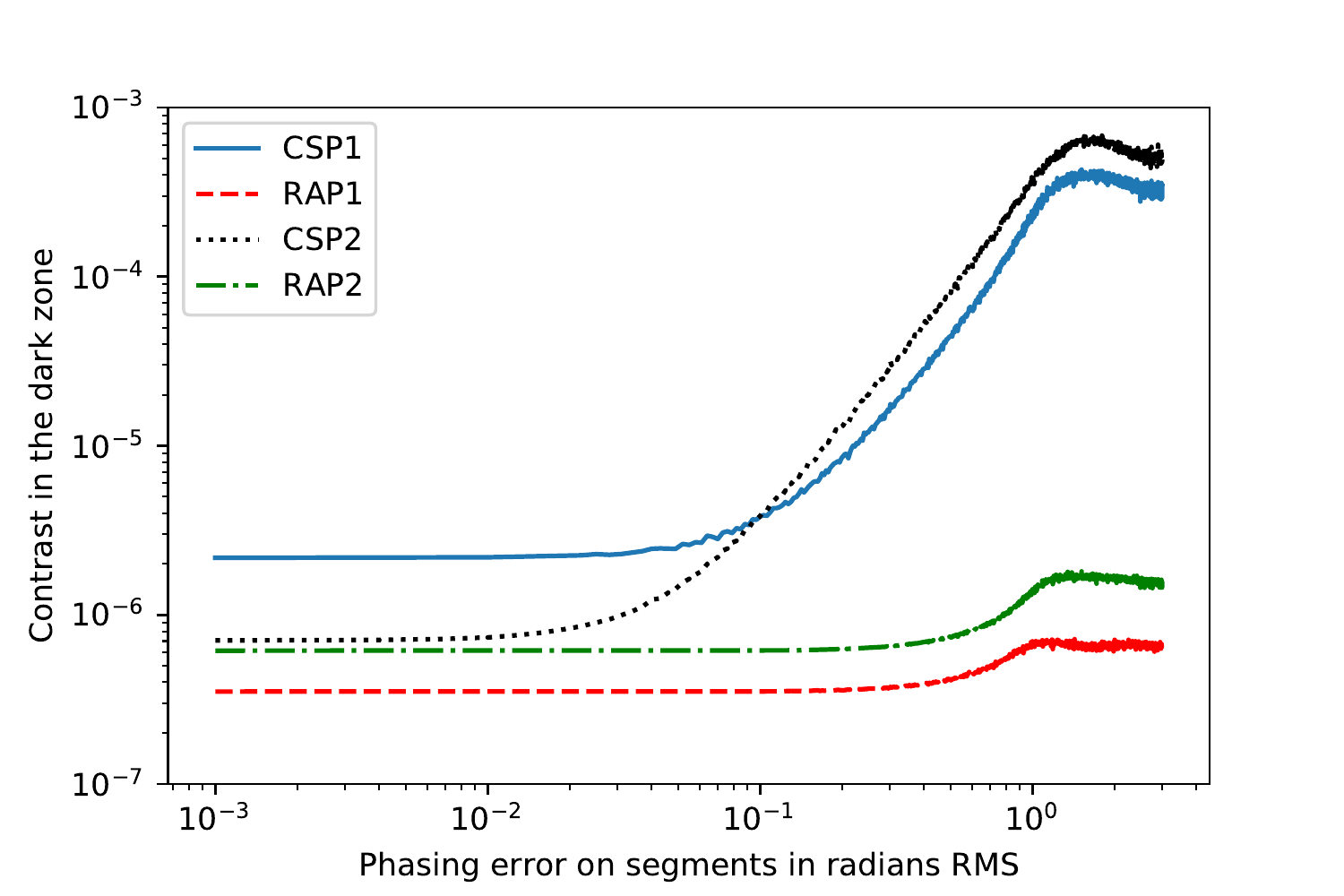}
   \end{tabular}
   \end{center}
   \caption[APP_HockeyCross] 
   { \label{fig:APP_HockeyCross} 
Evolution of the contrast with the phasing errors (blue) with the CSP1 design, (red) with the RAP1 design, (black) with the CSP2 design, and (green) with the RAP2 design. 1000 different error amplitudes are considered between 1 mrad RMS and 1 rad RMS. For each of them, 100 phasing errors are simulated and propagated, 100 PSFs normalized by their aberrated direct image peak are computed, and the average of the 100 resulting dark hole} contrasts is plotted.
   \end{figure}

We extended this first approach by computing the plot of Fig.\ref{fig:APP_HockeyCross}, which shows the contrast in the dark region as a function of the segment phasing-error amplitude for all four designs. Once again, 1000 different error amplitudes were considered between 1 mrad RMS to 3 rad RMS, and for each of them, 100 different random phasing errors were simulated and propagated through the coronagraph. The plot corresponds to the average of the 100 resulting contrasts. In the RAP cases, the phasing errors have no impact on the dark region contrasts, which remain almost constant with $3.5 \times 10^{-7}$ and $6.2 \times 10^{-7}$ for RAP1 and RAP2 until $\sim 1$ rad RMS, while the contrast deteriorates from a few dozen mrad for CSPs.

%-------------
%-------------

This robustness is caused by the absence of the second step, which would deteriorate the segment apodization optimization by modifying the low-order envelope $\left \Vert \widehat{Z}(\mathbf{u}) \right \Vert ^2$ of equation \ref{eq:PASTIS}. The dark zone contrast is then fully due to the apodized segment low-order envelope, and these results directly derive from Eq.\ref{eq:PASTIS}.

\subsection{Robustness to other errors}
\label{s:Comments on robustness to other errors2}

The primary mirror segmentation can cause other sources of errors, including tip-tilt phasing errors and missing segments. These are taken into account in this section.

\begin{figure}%[h]%*} [h]
   \begin{center}
   \begin{tabular}{c}
   \includegraphics[width=8.5cm]{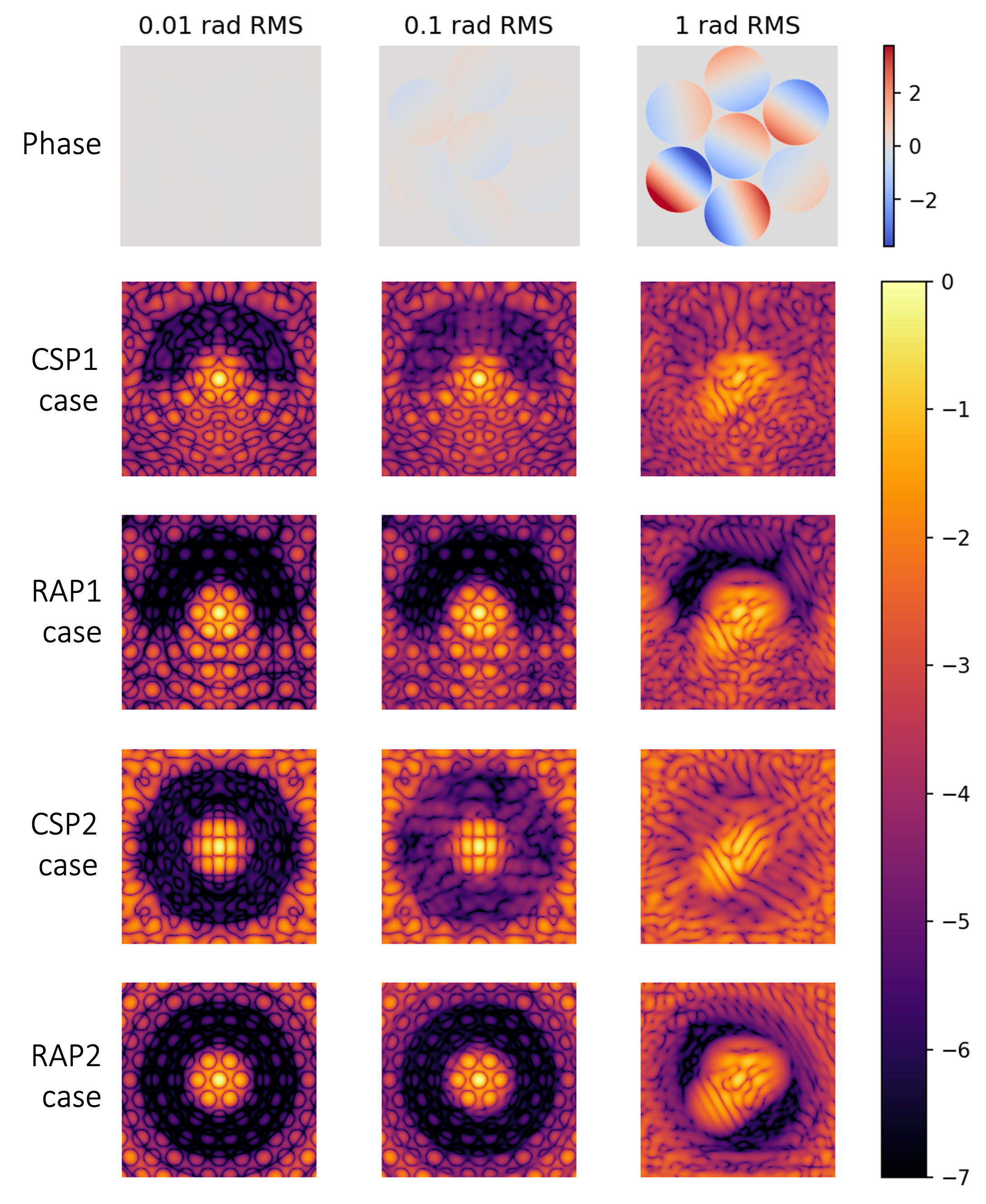}
   \end{tabular}
   \end{center}
   \caption[APP_TTErrorBudget] 
   { \label{fig:APP_TTErrorBudget} 
Robustness to tip-tilt phasing errors: (top) Tip-tilt phasing errors applied on the GMT-like pupil for three different amplitudes: 0.01 rad RMS, 0.1 rad RMS, and 1 rad RMS. (bottom) PSFs in logarithmic scale with the four APP designs CSP1, RAP1, CSP2, and RAP2 with the aberrations above.}
   \end{figure}

\begin{figure}
   \begin{center}
   \begin{tabular}{c}
   \includegraphics[width=8.5cm]{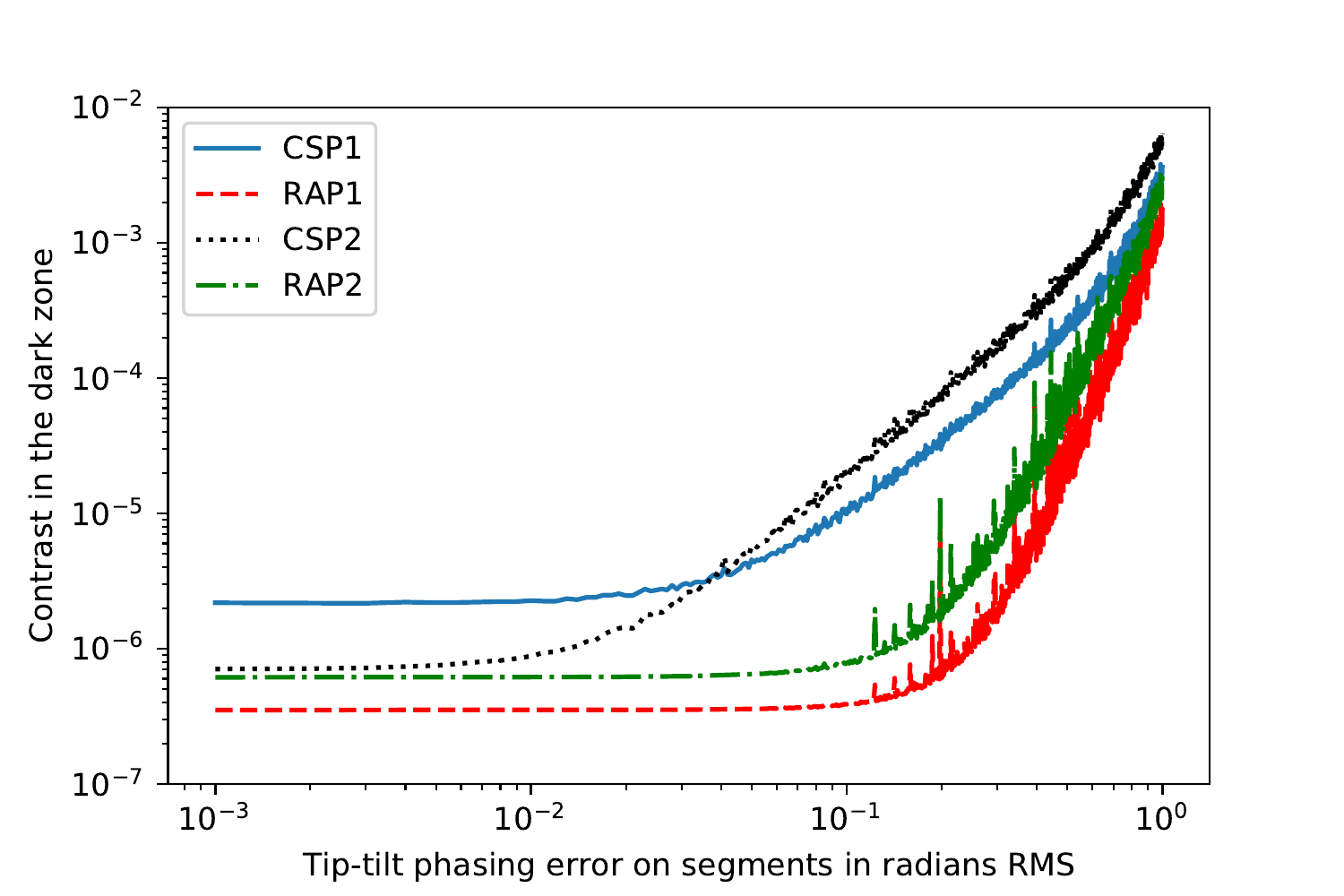}
   \end{tabular}
   \end{center}
   \caption[APP_TTHockeyCross] 
   { \label{fig:APP_TTHockeyCross} 
Evolution of the contrast with the tip-tilt phasing errors (blue) with the CSP1 design, (red) with the RAP1 design, (black) with the CSP2 design, and (green) with the RAP2 design. 1000 different error amplitudes are considered between 1 mrad RMS and 1 rad RMS. For each of them, 100 phasing errors are simulated and propagated, 100 PSFs normalized by their aberrated direct image peak are computed, and the average of the 100 resulting dark hole} contrasts is plotted.
   \end{figure}

First, in Fig.\ref{fig:APP_TTErrorBudget} segment-level tip and tilt errors are added to the primary mirror (top) with three different amplitudes ($0.01$, $0.1$, and $1$ rad RMS), and the resulting PSFs through the four designs are simulated (bottom). The RAP designs appear more robust to low aberrations, with almost no impact in the dark zone up to $0.1$ rad RMS of tip-tilt aberrations, which is confirmed by Fig.\ref{fig:APP_TTHockeyCross} for a wide range of aberrations. With both RAP designs, the contrast is constant in the dark zone until $0.1$ rad RMS before it increases, and it remains more robust than in the corresponding CSP cases. In addition, a contrast of $10^{-5}$ in the dark zone imposes constraints of $98$ versus $250$  mrad RMS in the CSP1 and RAP1 cases, and $69$ versus $142$ mrad RMS in the CSP2 and RAP2 cases, indicating an increased robustness for the RAP designs.

%-------------
%-------------

%-------------
%-------------

\begin{figure}%[h]%*} [h]
   \begin{center}
   \begin{tabular}{c}
   \includegraphics[width=8.5cm]{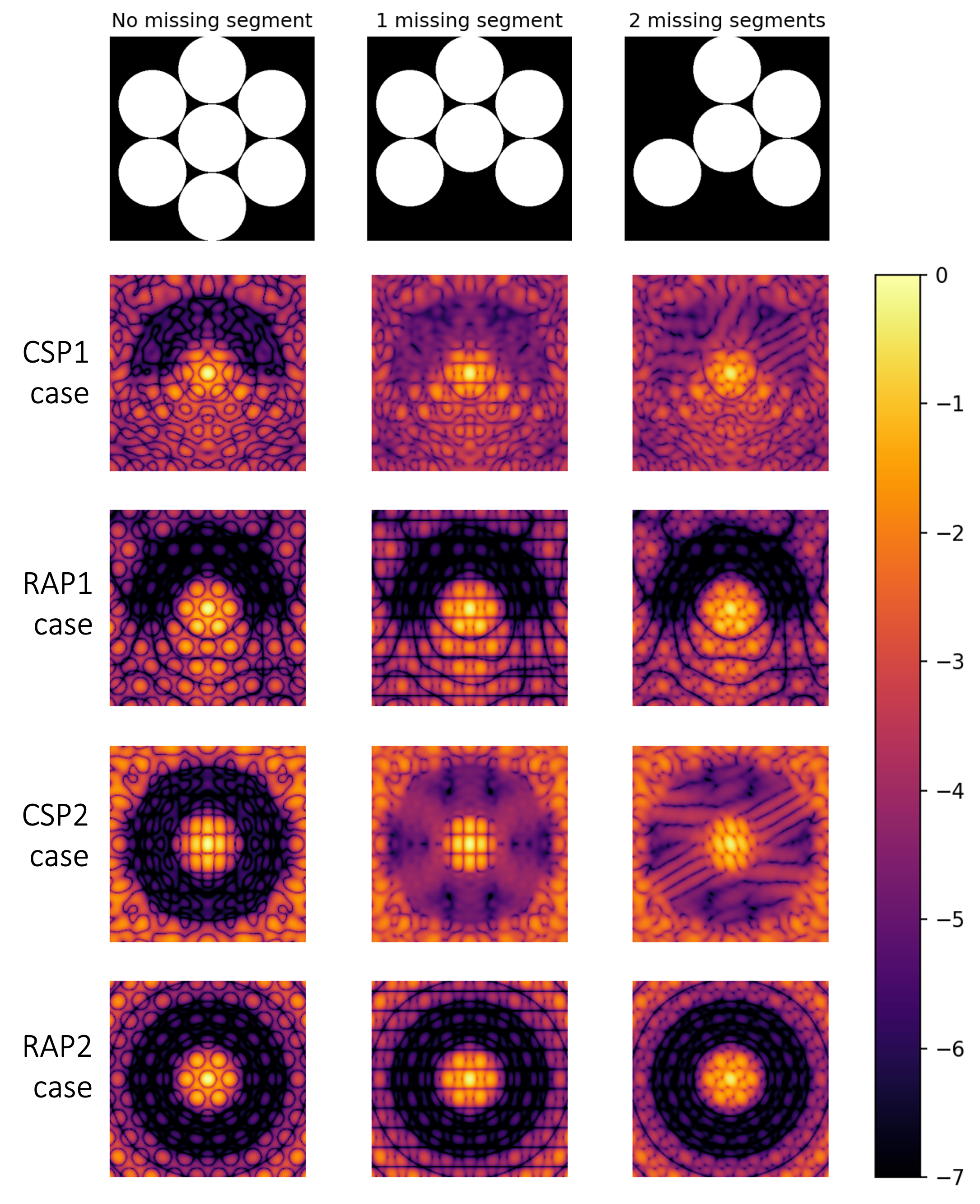}
   \end{tabular}
   \end{center}
   \caption[APP_MissingSegments] 
   { \label{fig:APP_MissingSegments} 
Robustness to missing segments: (top) Pupil configuration without a missing segment, with one missing segment, and with two missing segments. (bottom) PSFs in logarithmic scale with the four APP designs CSP1, RAP1, CSP2, and RAP2.}
   \end{figure}

As a second test, one to two of the seven segments of the primary mirror were removed. The impact of these amplitude aberrations on the four coronagraphic PSFs is illustrated in Fig.\ref{fig:APP_MissingSegments}. The contrasts of these PSFs are listed in Table~\ref{tab:APP_MissingSegments} and indicate that the RAP designs are far more robust to missing segments. Once again, this is due to the low-order segment envelope, which is not impacted at all by missing segments and maintains a contrast in the dark zone below $10^{-5.5}$. This result can be expected from Eq.\ref{eq:PASTIS} because in the RAP case, a missing segment does not impact the low-order envelope, but only the segment bases of the interference pattern.

%-------------
%-------------

\begin{table}%[!htb]
    \caption{Contrasts without and with one and two missing segments in the pupil.}
    \label{tab:APP_MissingSegments}
    \centering
    \begin{tabular}{l||c|c|c}
        & 0 mis. seg. & 1 mis. seg. & 2 mis. seg. \\
        \hline \hline
        CSP1 & $2.2 \times 10^{-6}$ & $2.4 \times 10^{-5}$ & $7.3 \times 10^{-5}$ \\
        \hline
        RAP1 & $3.5 \times 10^{-7}$ & $4.1 \times 10^{-7}$ & $4.8 \times 10^{-7}$ \\
        \hline \hline
        CSP2 & $7.1 \times 10^{-7}$ & $4.1 \times 10^{-5}$ & $1.2 \times 10^{-4}$ \\
        \hline
        RAP2 & $6.1 \times 10^{-7}$ & $7.3 \times 10^{-7}$ & $8.7 \times 10^{-7}$ \\
    \end{tabular}
\end{table}

\subsection{Conclusions for redundant APPs}
\label{s:Conclusions on redundant APPs}

RAP designs as redundant APPs appear highly robust: they are not impacted at all by piston-like phasing errors and missing segments, and they are less impacted by tip-tilt-like phasing errors than classical designs. This increased robustness compared to the RAP and APLC combination of section \ref{s:Numerical validation on a GMT-like aperture} is due to a more efficient first step (higher contrast of the envelope) and to the lack of the second step that would modify the redundance of the segment apodization and so the low-order envelope.

Compared to the APLC case of section \ref{s:Numerical validation on a GMT-like aperture}, the APP provides an access to smaller IWAs, but it is still limited to a few $\lambda/D$: In our study case, the contrast of $\sim 10^{-5.5}$ imposes an IWA of $1.83 \lambda/d$, that is, $5.5 \lambda/D$.

%%%%%%%%%%%%%%%%%%%%%%%%%%%%%%%%%%%%%%%%%%%%%%%%%%%%%%%%
\section{Conclusions}
\label{s:Conclusions}
%%%%%%%%%%%%%%%%%%%%%%%%%%%%%%%%%%%%%%%%%%%%%%%%%%%%%%%%

We have presented a method for designing RAPs, which are redundant apodizers that enable releasing the segment phasing constraints. We tested both two-step designs (segment-level optimization plus pupil-level optimization) with APLCs and one-step designs with purely segment-level APPs, which combine coherently in the detector for a GMT-like architecture, meaning seven circular segments, three along the pupil diameter.

In the APLC case, the two RAP designs proposed in this paper enable gaining a a factor of $\sim5$ to $\sim13$ on the segment piston alignment for a target contrast. To reach theses factors, the segment apodization is optimized in amplitude to dig its impact in the coronagraph dark zone. In the APP case, that is, with phase-segment apodizations, the studied RAP designs can remove any constraints on segment phasing for one-sided and two-sided dark zones. 

We also tested the robustness of the RAP designs to other segment-level errors in phase and amplitude: first with tip-tilt phasing errors that were not considered in the original designs, and then with missing segments. For APLC and APP, the RAP designs enabled reducing the tip-tilt constraints. For APLC, missing segments have a lower impact on RAP pupils than on classical pupils, and for APP, they have even almost no impact on the coronagraphic performance.

Two main drawbacks of the RAP design can be pointed out. First, the access to small inner working angles is limited, mainly for pupils made of many segments: The segment envelope cannot be optimized below $\sim 1.5-2 \lambda/d$, where $d$ is the segment diameter, and therefore it cannot impact the final coronagraphic PSF below $N \lambda/D$, where $N$ is the number of segment along the pupil diameter and $D$ is the pupil diameter. This phenomenon limits the direct application of RAPs to high-segment count mirrors such as the ELT or the TMT mirrors with which the segment envelope central peak fully covers the smallest angular separations (up to $\sim 30 \lambda/D$ for the ELT and up to $\sim 20 \lambda/D$ for the TMT). Redundant phase apodizers enable reducing the accessible inner working angle, but other options might be studied, such as phase-induced amplitude apodization (PIAA) coronagraphs that can reach $0.9 \lambda/d$ \citep{Guyon2014, Newman2016}. 

Second, mainly in the APLC case, the RAP designs lower the planet throughput, which becomes particularly an issue in the $10^{-5}$ envelope RAP design. In observation, this loss in throughput will affect the exposure time, and a trade-off should be found between the contrasts that are to be achieved in both steps regarding this throughput loss. However, for a two-sided dark zone, the redundant APP has a slightly higher throughput than the classical APP.

The issue of segment phasing is crucial because coronagraphs reach very high contrast with extreme sensitivity, degraded by phasing errors and segment vibrations, mainly for long exposure times. One of the most dramatic errors will concern the coming ground-based giant telescopes: a few segments of the ELT primary mirror will constantly be missing because they need to be replaced, which will generate diffraction effects up to $10^{-5}$ in contrast and will limite the access to faint planets. Our proposition is in line with these robustness issues, but at more modest angular separations.

The next step of this concept validation consists of an experimental setup of the RAP design, which requires a high-contrast testbed including a segmented mirror and an accessible apodizer plane. Several testbeds propose such configurations \citep{Mazoyer2019}, including the High-contrast imager for Complex Aperture Telescope (HiCAT) \citep{N'Diaye2013, N'Diaye2014, N'Diaye2015a, Leboulleux2016, Soummer2018, Laginja2020}, the High-Contrast High-Resolution Spectroscopy for Segmented telescopes Testbed (HCST) \citep{Jovanovic2018b, Llop-Sayson2019}, and the Segmented Pupil Experiment for Exoplanet Detection (SPEED) \citep{Janin-Potiron2018, Martinez2018, Beaulieu2018}.

RAPs can be seen as a static alternative to wavefront sensing and control loops that rely on a dynamic compensation of wavefront errors. Dynamic wavefront control requires adaptive components like one to two deformable mirrors and a wavefront sensor, while RAPs are directly integrated in the coronagraph and simplify the whole system in terms of hardware and software. A combination of RAP and dynamic wavefront control solutions can be considered for further studies, for instance, with the active compensation of aperture discontinuities (ACAD-OSM) \citep{Pueyo2013, Mazoyer2018, Mazoyer2018a}, to reduce the actuator strokes or to provide stability in complementary areas of the target dark zone. ACAD-OSM could focus on small angular separations and RAP farther from the star, which would release the constraints on the deformable mirrors: fewer actuators would be needed for the ACAD-OSM correction.

In addition, this method could be applied to another type of aberrations: island effects. First, the low-wind effect occurs for wind speeds lower than 1m/s that generate thermal gradients around the spiders, and consists of piston, tip, and tilt phase errors on the pupil petals defined by the spiders. This phenomenon limits high-contrast observations on VLT/SPHERE \citep{Sauvage2015, Sauvage2016, Milli2018} and at the Subaru telescope \citep{N'Diaye2018a, Bos2020} and is expected at the ELT. Second, post-adaptive optics petaling effects, consisting of petal-level pistons falsely reconstructed by the adaptive optics system when the spiders are thicker than the atmosphere Fried parameter, are also expected at the ELT \citep{Bertrou-Cantou2020, Schwartz2018}. The redundance of the petal pattern can be taken advantage of to design RAPs on the six ELT petals that are identical to each other despite the rotation, to generate coronagraphs that are robust to petal-level pistons, tips, and tilts, that is, to both low-wind effect and post-adaptive optics petaling. This study is the purpose of a second paper.

\begin{acknowledgements}
This project is funded by the European Research Council (ERC) under the European Union's Horizon 2020 research and innovation programme (grant agreement n$^\circ$\,866001). We also thank the referee for their useful comments which improved this article a lot.
\end{acknowledgements}

\bibliographystyle{aa}
\bibliography{bib}

\end{document}